\documentclass[a4paper,10pt]{article}

\usepackage{amssymb}
\usepackage{amsmath}
\usepackage[usenames,dvipsnames]{color}
\usepackage{hyperref}
\usepackage[left=.8in,right=.8in,top=.8in,bottom=.8in]{geometry}                  

\hypersetup{
    colorlinks=true,         
    linkcolor=blue,          
    citecolor=red,        
    urlcolor=Violet             
}

\newcommand{\mean}[1]{\ensuremath{\lf\langle #1 \rt\rangle }}
\newcommand{\diby}[2]{\ensuremath{\frac{\delta #1}{\delta #2}}}

\newcommand{\order}[1]{\ensuremath{\mathcal{O}(#1)}}

\def\be{\begin{equation}}
\def\ee{\end{equation}}
\def\bea{\begin{eqnarray}}
\def\eea{\end{eqnarray}}

\def\lf {\ensuremath{\left}}
\def\rt {\ensuremath{\right}}

\title{Poincar\'e invariance and asymptotic flatness in Shape Dynamics}
\author{Henrique Gomes}
\author{\bf Henrique Gomes\footnote{\href{mailto:gomes.ha@gmail.com}{gomes.ha@gmail.com}}\\\it University of California at Davis\\ \it One Shields Avenue Davis, CA, 95616, USA }

\begin{document}

\maketitle

\begin{abstract}Shape Dynamics is a theory of gravity that waives refoliation invariance in favor of spatial Weyl invariance. It is a canonical theory, constructed from a Hamiltonian, 3+1 perspective. One of the main deficits of Shape Dynamics is that its Hamiltonian is only implicitly constructed as a functional of the phase space variables. In this paper, I write down the equations of motion for Shape Dynamics to show that over a curve in phase space representing a Minkowski spacetime, Shape Dynamics possesses Poincar\'e symmetry for appropriate boundary conditions. The proper treatment of such boundary conditions leads us to completely formulate Shape Dynamics for open manifolds in the asymptotically flat case. We study the charges arising in this case and find a new definition of total energy, which is completely invariant under spatial Weyl transformations close to the boundary. We then use the equations of motion once again to find a non-trivial solution of Shape Dynamics, consisting of a flat static Universe with a point-like mass at the center. We calculate its energy through the new formula and rederive the usual Schwarzschild mass.
\end{abstract}

\tableofcontents
\section{Introduction}

Shape Dynamics is a theory of gravity. It is a canonical theory, constructed from a Hamiltonian, 3+1 perspective. Classically, it differs little from a gauge-fixation of the canonical form of General relativity \cite{SD:FAQ}. To be more precise, a gauge-fixed version of Shape Dynamics is identical to General Relativity in a particular gauge of its own - the so called CMC gauge. In the quantum realm however, the differences might be great. That is because Shape Dynamics does not possess refoliation invariance. Instead, it has complete spatial Weyl invariance. From an effective field theory point of view, this would already imply a vastly different scenario: gauge-invariant terms in Shape Dynamics are spatially conformal-diffeomorphism invariant, as opposed to invariant under space-time diffeomorphisms. 

 Regarding a classical distinction, it  can be shown that there are global solutions of general relativity that cannot be put completely into a Shape Dynamics formulation, just as there might be solutions of Shape Dynamics that might not have a GR counterpart. General relativity is experimentally very tightly constrained on one hand,  but on the other it gives rise to singularities and is not understood as a quantum theory. It is interesting therefore to  have an example of a theory that differs from general relativity, but not too much. This provides a good motivation to study  classical aspects of Shape Dynamics.
 
 Shape Dynamics is grounded upon certain uniqueness and existence results for a particular type of initial value formulation of GR \cite{York:1972sj}. Its Hamiltonian, although guaranteed to exist and be unique under a generous set of circumstances, is given  only implicitly as a solution of a second order differential equation over each point in phase space. Despite this abstract quality, it has already served to yield much insight into Weyl invariance on GR \cite{SD1, SD:LT}, holographic renormalization \cite{HJSD}, symmetry doubling in GR \cite{SymDoub}, AdS/CFT \cite{Flavio and Sean}, and more general symmetry trading \cite{SymDoub, SD:FAQ}. 

In this paper we aim to take Shape Dynamics into two somewhat different directions.  First, we would like to study particular solutions of Shape Dynamics, and not just general theoretical features. 
For this it is not enough to formulate the theory implicitly, by merely finding its Hamiltonian. We must in fact find a manageable way to write the equations of motion for Shape Dynamics, which can then be studied over each solution separately. We do this in section \ref{sec:eom}. Using the equations of motion, we will see the emergence of Poincar\'e invariance from  Shape Dynamics, which arises naturally over a curve in phase space that represents a Minkowski spacetime. 
Given that the field invariance present in Shape Dynamics (spatial dilatation and spatial diffeomorphism invariance) is vastly different than that present in ADM (on shell spacetime diffeomorphisms), the  formal equivalence of the two theories has seemed puzzling in some respect. We hope this result will go some ways towards elucidating this puzzle.  

The study of the Minkowski analogue in Shape Dynamics will naturally bring us to focus our attention on the second purpose of this paper: the formulation of Shape Dynamics for open spatial manifolds.  In the open manifold case Shape Dynamics possesses simpler equations of motion than for the closed manifold setting, in which   Shape Dynamics has so far concentrated. On the other hand closed topology is simpler in that it does not require us to deal with boundary conditions, whereas the open spatial manifold case involves this complication. 

 These considerations require us to formulate and understand the construction of Shape Dynamics in the presence of boundaries and, in particular, in the asymptotically flat case.  Limiting our field space to obey one  set of natural  boundary conditions will produce, for a curve of Minkowski initial data,  a theory with full Poincar\'e symmetry.\footnote{ A note on nomenclature: we will distinguish ``invariance", as in diffeomorphism invariance, from ``symmetry", as in Poincar\'e symmetry. The first is a covariance condition, under which the action is invariant. This should be a property holding throughout phase space. The second is a transformation which leaves a given point (or a curve) in phase space fixed.} That Poincar\'e symmetry and its charges arise is an accident of the particular lapse propagators in Shape Dynamics.
 As an indirect consequence of this ``accident", in an explicit calculation of the charges arising in the boundary formulation we find an extra term to the mass charge of ADM. This charge is seen to correct the ADM mass so that the mass for asymptotically flat Shape Dynamics is preserved under asymptotic spatial Weyl transformations. 

Lastly, we will use the equations of motion of Shape Dynamics once more to find a solution  for static phase space data, consisting of a flat metric and vanishing momenta,  with a point mass at the center of the Universe. We show that we can find a unique solution for Shape Dynamics in this instance. The lapse is uniquely fixed to be the one for Schwarzschild solution in isotropic coordinates, and if one reconstructs the spatial metric one obtains the full isotropic Schwarzschild solution. The Shape Dynamics definition of mass  is used for this solution and we still obtain in a simple manner the correct Schwarzschild mass.

We start in  section \ref{sec:LT} with the ADM Hamiltonian and its constraint algebra. We then briefly present the construction of Shape Dynamics in the closed manifold case, for those unacquainted. In section \ref{sec:ps_red} we study  phase space reduction in the presence of boundaries  in its generality. This is a rather abstract and technical section, and can be skipped at a first reading, in favor of the concrete results of section \ref{sec:Boundaries}.   Section \ref{sec:ps_red} consists in defining the leftover total Hamiltonian, and its consistency with the reduction procedure. We see that the only instance in which we are left with a non-trivial evolution generator for Shape Dynamics is if the boundary conditions on the lapse allow lapses that propagate maximal slicing on the boundary. We end the section by writing out the equations of motion of Shape Dynamics for the open manifold case.  Section \ref{sec:Boundaries} is where we concentrate our more concrete results. We look at the solution over Minkowski-like initial data and find both 3-dimensional conformal symmetry and 4-dimensional Poincar\'e symmetry. These symmetries, however, don't close on themselves, and would thus prompt a tower of constraints. By investigating the proper formulation of the boundary generators we find the root of the inconsistency: in order to ensure finiteness and differentiability of the constraints the two symmetries cannot be simultaneously well-defined. The charges resulting from these investigations are calculated, and seen to coincide with ADM, except for the energy. The new energy is shown to be Weyl invariant.  Finally we find a non-trivial ansatz and solution for Shape Dynamics which reproduces the Schwarzschild solution in preferred isotropic coordinates. We summarize our results in section \ref{sec:conclusions}.  In appendix \ref{appendix:LFE_CMC} we comment on how each set of symmetries excludes the other in the closed, CMC slicing case.  In appendix \ref{app:Boundaries} we concentrate some of the auxiliary calculations for the boundary variations, and in appendix \ref{appendix:central} we show that central terms don't prevent the constraints in Shape Dynamics with our appropriate boundary conditions from forming a Poincar\'e algebra. 


\section{From ADM to Shape Dynamics}\label{sec:LT}
\subsection{The ADM constraints}

 Let us begin by writing out the constraints of canonical GR in its 3+1 ADM form: 
\begin{eqnarray}
\label{equ:scalar constraint}S(x):= \frac{G_{abcd}\pi^{ab}\pi^{cd}}{\sqrt g}(x)-R(x)\sqrt g(x)=0\\
\label{equ:momentum constraint} H_a:={\pi^{a}_b}_{;a}=0
\end{eqnarray}
The scalar constraint \eqref{equ:scalar constraint} generates on-shell refoliations of spacetime, while the momentum constraint generates foliation preserving diffeomorphisms.

The algebra of the constraints is 
\begin{subequations}\label{equ:ADM_algebra}
\begin{eqnarray}
\{S(N_1),S(N_2)\}&=&H_b(g^{ab}(N_1\partial_a N_2-N_2\partial_a N_1))\label{equ:ADM_alg:SS}\\
\{S(N),H^a(\xi_a)\}&=&-S(\mathcal{L}_{\vec\xi} N)\\
\{H_a(\xi^a),H_ b(\eta^b)\}&=& H_a([\vec\xi,\vec\eta]^a)
\end{eqnarray}
\end{subequations}
where we use the notation for smearing $S(N)=\int d^3 x N(x)S(x)$ and  $H^a(\xi_a)= \int d^3 x H^a(x)\xi_a(x) $, $N\in C^\infty(M)$ and $\xi^a\in \Gamma^\infty(TM)$ is a smooth vector field.  

The algebra of constraints above can be put in a general form 
\be\{H^\mu(\xi_\mu),H^\nu(\eta_\nu)\}=H([\xi, \eta]_{\mbox{\tiny SD}})
\ee
where $H_0=S$, $\xi^0=N$ in the above, and $[\xi, \eta]_{\mbox{\tiny SD}}$ is the so  called \emph{hypersurface deformation algebra of vector fields}, and stands for the commutator of vector fields orthogonally  decomposed along the hypersurface.

\subsection{Shape Dynamics}
Here we will briefly review  the construction of Shape Dynamics as a theory equivalent to ADM gravity on a compact Cauchy surface $\Sigma$ without boundary. For details see \cite{SD:LT}.

 We start with the standard ADM phase space $\Gamma_{ADM}=\{(g,\pi):g\in \mathrm{Riem},\pi\in T_g^*(\mathrm{Riem})\}$, where $\mathrm{Riem}$ denotes the set of Riemannian metrics on the 3-manifold $\Sigma$ with the usual first class ADM constraints. 
Now we extend the ADM phase space with the phase space of a scalar field $\phi(x)$ and its canonically conjugate momentum density $\pi_\phi(x)$. The trivial embedding of the original system into this extended phase space introduces an additional first class constraint ${Q}(x)=\pi_\phi(x)\approx 0$. 

To obtain a non-trivial embedding of the theory on the extended phase space we use the canonical transformation $t_\phi$ generated by the generating functional 
\be\label{equ:can_gen}F=\int d^3x\left(g_{ab}e^{4\hat \phi}\Pi^{ab}+\phi\Pi_\phi\right),\ee
 with $\hat \phi(x):=\phi(x)-\frac 1 6 \ln\langle e^{6\phi}\rangle_g$ where the mean $\mean{\cdot}$ is defined as $\langle f\rangle_g:=\frac 1 {V_g} \int d^3x\sqrt{|g|} f(x)$ and the 3-volume is $V_g:=\int d^3x\sqrt{|g|}$. The triangle brackets - which we call ``means" -  and the volumes technically appear as a manner to force the appearance of a non-zero global Hamiltonian in Shape Dynamics. We will have more to say about this later. The transformations effected by the generating functional on the canonical variables are:  
\be\label{equ:can_transf}\begin{array}{rcl}
    t_{\hat\phi} g_{ab}= e^{4 \hat \phi(x)} g_{ab}&,& t_{\hat\phi}\pi^{ab}=e^{-4 \hat \phi} \left(\pi^{ab}-\frac 1 3 \langle \pi\rangle_g \left(1-e^{6\hat \phi}\right)g^{ab}\sqrt{|g|}\right)\\
    ~&~&~\\
    t_{\hat\phi} \pi_\phi=\pi_\phi-4\left(\pi(x)-\sqrt{g}(x)\langle\pi\rangle_g\right)&,& t_{\hat\phi} \phi= \phi. 
\end{array}\ee
 At this point we obtain the first class set of constraints
\begin{eqnarray*}
  t_\phi S(x)= S(t_{\hat\phi}g_{ab}(x),~ t_{\hat\phi} \pi^{ab}(x)),~
   t_\phi H(x)= H(t_{\hat\phi} g_{ab}(x),~ t_{\hat\phi} \pi^{ab}(x))\\
    ~&~& ~\\
     \mbox{and}~~~
   t_{\hat\phi} Q(x)=\pi_\phi(x)-4(\pi-\langle\pi\rangle\sqrt g)(x)
\end{eqnarray*} 
 We call this system \emph{the Linking Theory}.

 As the last step to obtain Shape Dynamics, we perform a phase space reduction to go back to the original phase space consisting of just the variables $(g,\pi)$. The gauge fixing condition we impose is $\pi_\phi(x)=0$. For the phase space reduction to be well-defined,  the gauge orbit must intersect the constraint surface exactly once.

At the constraint surface $\pi_\phi=0$, the constraint $ S(t_{\hat\phi}g_{ab}(x), t_{\hat\phi} \pi^{ab}(x))=0$ is:  
\be\label{equ:non_homogeneous LY} t_\phi\left( \frac{S(x)}{\sqrt g}\right)=e^{-4\hat\phi}\left(R-8(\nabla^2\hat\phi+\phi_{,a}\phi^{,a})\right)-\frac{e^{-12\hat\phi}}{g}\left(\pi^{ab}\pi_{ab}-\frac{1}{2}\pi^2\right)=\mean {\mbox {l.h.s}}
\ee
Where $\hat\Omega=e^{\hat\phi}$, and $\mean {\mbox {l.h.s}} $ means that the left hand side of equation \eqref{equ:non_homogeneous LY} has to be a spatial constant. Generically this equation has a one parameter family of solutions, one for each constant value of $\mean {\mbox {l.h.s}} $. It is only when one restricts the conformal factor $\Omega$ to be also volume-preserving, that is $V_{\Omega^4 g}=V_g$, that we obtain uniqueness \cite{SD1}.

The effect of using the volume preserving Weyl transformations then is that one does not get exactly the Lichnerowicz-York equation:
\be\label{equ:LY}e^{4\phi}\left(R-8(\nabla^2\phi+\phi_{,a}\phi^{,a})\right)-\frac{e^{-12\phi}}{g}\left(\pi^{ab}\pi_{ab}-\frac{1}{2}\pi^2\right)=0\ee 
 which says that the left hand side of \eqref{equ:non_homogeneous LY} is zero (with the additional requirement that $\phi$ be volume preserving, $\phi\rightarrow \hat\phi$). But a non-homogeneous form where the left hand side of \eqref{equ:non_homogeneous LY} has to be equal to a spatial constant. This is the form of the equation we have to solve for Shape Dynamics in case the manifold is compact without boundary, and it allows the appearance of a global Hamiltonian, as we have mentioned before (see also  appendix \ref{appendix:LFE_CMC}).  It corresponds to a constant mean curvature foliation of ADM \cite{SD1}, i.e. $\pi=\mean{\pi}\sqrt{g}$. Removing the total volume-preserving restriction, the generated conformal transformations would be related not to CMC gauge, but to maximal slicing. However, in the closed manifold case using the unrestricted conformal transformations would yield an identically vanishing Hamiltonian upon phase space reduction of the Linking theory.

 The end result is the following set of first class constraints, together with the usual Poisson bracket for the metric variables, defining Shape Dynamics
\begin{equation}\label{equ:pureSDconstraints}
 \begin{array}{rcl}
   H_{\mbox{\tiny{SD}}}&=&\mean{t_{\phi_o}\left(\frac{S(x)}{\sqrt g}\right)}\\
    ~&~& ~\\
   H^a(\xi_a)&=&\int d^3x( \pi^{ab}\mathcal{L}_\xi g_{ab})\\
    ~&~& ~\\
   D(\rho)&=&\int d^3x \rho\left(\pi-\langle \pi\rangle\sqrt{|g|}\right),
 \end{array}
\end{equation}
Here $\rho$ is the parameter of the gauge transformation that will be effected by the constraint (i.e. it is the Lagrange multiplier of that constraint), $\phi_o$ is the solution of \eqref{equ:non_homogeneous LY}.

For maximal slicing $\pi=0$, Shape Dynamics for closed manifolds has a zero Hamiltonian, and is equivalent to ADM only as a theory of initial data. 
 But for open manifolds there are boundary terms, and  it is unclear what one would obtain using the above construction. We will discuss the construction in section \ref{sec:ps_red}. 
As it happens, in case our manifold has a boundary, we can implement maximal slicing and still have an evolution Hamiltonian. For maximal slicing, the required conformal transformations are general, i.e. not necessarily volume preserving. This simplifies some of the equations, for instance the equation defining the conformal factor, 
\be t_\phi S(x)=e^{2\phi}\sqrt g\left(R-8(\nabla^2\phi+\phi_{,a}\phi^{,a})\right)-\frac{e^{-6\phi}}{\sqrt g}\left(\pi^{ab}\pi_{ab}-\frac{1}{2}\pi^2\right)=0\ee 
which is  equivalent to the original Lichnerowicz-York equation \eqref{equ:LY}, \cite{York:1972sj}. In the maximal slicing case, we still have \begin{equation}\label{equ:maximal_SDconstraints}
 \begin{array}{rcl}
   H^a(\xi_a)&=&\int d^3x( \pi^{ab}\mathcal{L}_\xi g_{ab})\\
   ~&~& ~\\
   D(\rho)&=&\int d^3x \rho\pi,
 \end{array}
\end{equation}
The construction of the evolution Hamiltonian however, will be the subject of sections \ref{sec:ps_red} and \ref{sec:Boundaries}.

\subsubsection{Remarks on the Shape Dynamics Hamiltonian and the propagating lapse.}  

 That a one parameter family of solutions for \eqref{equ:non_homogeneous LY} exists  is also indicated by a one-dimensional kernel for the propagation of the gauge-fixing $\pi_\phi=0$. This kernel is the kernel of the Poisson bracket: $\{\pi_\phi(x),t_{\hat\phi} S(N)\}$ \cite{SD:LT}. We will often regard the equation obtained from this as an operator on functions $N(x)$.  
The equation is given by:
\be\label{equ:tphi_LFE_abstract}\{t_{\hat\phi} S(N), \pi_\phi(x)\}=\diby{t_{\hat\phi} S}{\phi}\cdot N
\ee

Thus Shape Dynamics leaves out a generator with the kernel of \eqref{equ:tphi_LFE_abstract} as a constraint which is not gauge-fixed by $\pi_\phi=0$. We usually call this particular choice of lapse, the one that propagates our gauge-fixing,  the solution to the lapse fixing equation: \emph{the propagating lapse, $N_o$}. This is not great nomenclature, but by force of habit we will resort to it many times in this paper. 
Using the propagating lapse and the LY factor $\phi_o$, we can alternatively write the Shape Dynamics propagating Hamiltonian as $t_{\phi_o}S(N_o)$. This form of the Hamiltonian, as we will see, will be  more useful in the presence of boundaries. 

If we do not impose that the Weyl transformation be volume-preserving (which is equivalent to requiring maximal slicing) the only solution to \eqref{equ:tphi_LFE_abstract} on a closed manifold  is $N_o=0$. And thus Using the $t_{\phi_o}S(N_o)$ form of the Hamiltonian, it is also easy to see that  there is  no Hamiltonian generating evolution. 

 Equation \eqref{equ:tphi_LFE_abstract} has a ``secondary status" to the  defining equation \eqref{equ:non_homogeneous LY}, since \eqref{equ:tphi_LFE_abstract} is a map tangential to $t_{\hat\phi} S(x)$ along $\phi$, and thus a unique solution to \eqref{equ:non_homogeneous LY} implies a unique solution to \eqref{equ:tphi_LFE_abstract}. That is, if there is a unique solution to equation \eqref{equ:non_homogeneous LY}, it will automatically project any lapse onto the preferred lapse $N_o$. Thus we obtain, upon phase space reduction, $t_{\hat{\phi}}S(N)\mapsto t_{{\phi_o}}S(N_o)$, which is the left over evolution Hamiltonian in Shape Dynamics for closed manifolds. 
 
 This is true both in the closed case and in the open case, and prompts one in many instances not to worry about the lapse fixing equation. However, in the present work we will need to pay it its due attention, because as we will see,  we are technically unable to obtain the equations of motion of Shape Dynamics directly from  the implicit function $\phi_o$.

\section{Phase space reduction in the presence of boundaries}\label{sec:ps_red}

\subsection{Boundary terms}

In the presence of a spatial boundary, the variational derivatives of the constraints are not well-defined \cite{RT}. This happens because one obtains terms from integration by parts that contain variations of the fields and its derivatives on the boundary, and these variations do not necessarily vanish. The antidote for this lack of differentiability is to add appropriate boundary terms to the constraints, whose variations should exactly cancel the boundary elements {of the variation of the bulk Hamiltonian}.
It is through the addition of such boundary terms that one defines energy and momentum for generally covariant theories,  since on-shell the constraints vanish and only the boundary terms survive.

More explicitly, suppose that $H^\mu$ is the constraint whose boundary variation we have to cancel, and $\xi_\mu$ is its appropriate smearing. Then we define (following \cite{Carlip}):
\be \bar H^\mu(\xi_\mu)=H^\mu(\xi_\mu)+B^\mu(\xi_\mu)=\int_\Sigma d^3 x \xi_\mu H^\mu+\int_{\partial\Sigma}d^2y \xi_\mu B^\mu
\ee
 where the variation of the boundary term cancels the boundary terms of the variations on the  bulk. \footnote{ This formalism is not complete for the consideration of the "cosmological"  setting, as, alongside \cite{RT, Hanson, BO} it takes the Hamiltonian ADM setting as a starting point and thus does not require the inclusion of  boundaries in time.  } 
In the Linking theory, we can straightforwardly find the terms that the  $B^\mu(\xi_\mu)$ ought to cancel.  

\subsection{The evolution Hamiltonian for Shape Dynamics in the presence of boundaries}


To render the constraints in the Linking Theory differentiable we must add to them boundary terms, as explained in the previous section. In particular, we must add a boundary term to the scalar constraint $t_\phi S(x) \rightarrow t_\phi(S(N)+B(N))$. 
By solving the LY equation \eqref{equ:LY}, which is just $t_\phi S(x)=0$, with specified  boundary conditions on $\phi$,  one is still left with the boundary generator $t_\phi(S(N)+B(N))\mapsto t_{\phi_o}B(N_o)$, where $\phi_o$ is the solution for $t_\phi S(x)=0$ with the given boundary conditions, and $N_o$ is a solution of the lapse fixing equation \eqref{equ:tphi_LFE_abstract} with its own boundary conditions.   In other words, by solving $t_\phi S(x)=0$ there is no ``left over" constraint at the bulk. However,  we still get the Hamiltonian generator at the boundary,  $t_\phi(S(N)+B(N))\mapsto t_{\phi_o}B(N_o)$.\footnote{In the closed case, if one uses the non-restricted conformal transformation (the unhatted $\phi$) - which is related to maximal slicing - one obtains a trivial Hamiltonian. This is the reason in that case one uses the volume-preserving condition.}

Regarding the propagating lapse, by the usual (min-max) arguments for invertibility of the equation \eqref{equ:tphi_LFE_abstract},  seen as a differential operator this equation is invertible for functions \emph{with compact support}.\footnote{In the generic case where $\pi^{ab}\neq 0$, otherwise we have more degeneracy.} It forms an invertible Dirac bracket for bulk functions, but allows a non-trivial Hamiltonian at the boundary. Given boundary conditions that only allow functions with compact support, the only propagating lapse is the zero lapse. The solution should be non-zero if there is to be any evolution left, and this requires that the boundary conditions allow for a non-zero solution of the lapse fixing equation, i.e. a propagating lapse, for the homogeneous boundary geometry (e.g. flat boundary geometry).


We reach the conclusion that Shape Dynamics on open manifolds and  maximal slicing possesses an evolution generator of dynamics \emph{only} at the boundary, and is in that sense holographic. Of course, this Hamiltonian is non-local, containing integrals over the bulk in the implicit solution $\phi_o$, and this allows us to recover a non-trivial time evolution in the bulk as well.

Let us briefly mention that the formulation $t_\phi S(x)=0$ of the constraint in the Linking theory lends itself more naturally to phase space reduction than for instance $\phi(x)-\phi_o(x) =0$, since it is harder to see what is to be the leftover Hamiltonian. One way this could be done would be to formulate appropriate restrictions on the conformal factor $\phi$ as additional, reducible constraints, and then insert the solution of the LY equation therein. We will not follow this route.

\subsection{Algebra of constraints}

 To calculate the algebra of the symmetries, we will need the algebra for the different generators. These can be obtained easily, by first writing down the Hamiltonian for the Linking Theory with full symmetry trading (i.e. in maximal slicing) in a closed space,  as
\be t_\phi\mathcal{H}(N, \xi^a, \rho)=t_{\phi}S(N)+H_a(\xi^a)+4\pi \mathcal{L}_\xi\phi+(\pi_\phi-4\pi)(\rho)
\ee
calculating the brackets and performing phase space reduction.  The total Hamiltonian for Shape Dynamics is  given by:
\be \mathcal{H}(N_o^{(i)}, \xi^a, \rho)=t_{\phi_o}(S(N_o^{(i)} ))+H_a(\xi^a)+4\pi(\rho)
\ee
where $\rho$ is the gauge parameter of the Weyl transformation.

 The algebra of constraints of SD emerging from this is easily calculated with the canonical transformation properties of the transformation $t_\phi$.\footnote{There is a slightly more complicated story here than we show. Phase space reduction would apparently produce the constraint $\pi(\rho+\mathcal{L}_\xi\phi_o)$. However, this can be checked not to alter the first class properties of the constraints and thus can be absorbed in $\rho$,  the Lagrange multiplier of $\pi$. The same occurs when using volume-preserving Weyl transformations.}
\begin{multline}\label{equ:SD_algebra} [\mathcal{H}(N_o^{(i)}, \xi^a, \rho), \mathcal{H}(N_o^{(i')}, \xi'^{a'}, \rho')]=\\
\mathcal{H}\left((\xi'^{a'}{N^{(i)}_o}_{,a'}-\xi^{a}{N^{(i')}_o}_{,a}), ~(g^{cd}({N^{(i)}_o}_{,c}N_o^{(i')}-{N^{(i')}_o}_{,c}N_o^{(i)})+ [\mathbf{\xi},\xi']^d),~ \xi'^{a'}\rho_{,a'}-\xi^{a}\rho'_{,a}\right)
\end{multline}
where only the solutions of the lapse fixing equation are allowed in the smearing. Upon the addition of boundary terms, all we have to do is to rewrite the projection of the constraint by the projection of the constraint with its boundary term, as in $t_{\phi_o}S(N_o)\mapsto t_{\phi_o}(S+B)(N_o)=t_{\phi_o}B(N_o)$. Furthermore,   the algebra may change by the  central terms, which we address in the appendix since they don't arise for the boundary conditions we will be exploring.

\subsection{Equations of Motion}\label{sec:eom}

\subsubsection{Preliminaries}

Now that we have made clear what we mean by the construction of Shape Dynamics on a manifold with boundary, we can write the equations of motion for the maximal slicing case. We will accomplish this by exploring one of the properties of the Dirac brackets occurring in our phase space reduction. Let us start by illustrating the procedure with the closed manifold case.

 In the closed manifold case what we want to ultimately calculate is the Poisson bracket $\{g_{ab}, t_{\hat\phi_o}S(N_o) \}$. We have that $t_{\hat\phi}S(x)-t_{\hat\phi}S(N_o)\sqrt g$ is second class with respect to $\pi_\phi$, and under reduction $t_\phi(\frac{ S(x)}{\sqrt g})\mapsto t_{\hat\phi_o}S(N_o)$, where $N_o$ is such that $\{t_{\hat\phi}S(N_o),\pi_\phi\}=0$.  By using the Dirac bracket, we can impose the reduction before or after calculating the brackets \cite{HT}. The Dirac bracket is defined as:
 \be\label{equ:Dirac_closed}
 \{\cdot, \cdot\}_{\mbox{\tiny{DB}}}=\{\cdot ,\cdot\}-
 \{~\cdot~,\pi_\phi\}\Delta^{-1}\{ t_{\hat\phi}S-t_{\hat\phi}S(N_o)\sqrt g,~\cdot~\}-\{~\cdot~,t_{\hat\phi}S-t_{\hat\phi}S(N_o)\sqrt g\}\Delta^{-1}\{\pi_\phi ,~\cdot~\}
 \ee
 where $\Delta= \{\pi_\phi, t_{\hat\phi}S-t_{\hat\phi}S(N_o)\sqrt g \}= \{\pi_\phi, t_{\hat\phi}S \}$. Thus
\be\label{equ:eom_closed}
\{g_{ab}, t_{\hat\phi}S(N) \}_{\mbox{\tiny{DB}}}=\{g_{ab}, t_{\hat\phi}S(N) \}-\{g_{ab}, (t_{\hat\phi}S-t_{\hat\phi}S(N_o)\sqrt g)(N) \}
\ee 
Upon reduction we obtain $\{g_{ab}, t_{\hat\phi}S(N) \}_{\mbox{\tiny{DB}}}=\{g_{ab}, t_{\hat\phi_o}S(N_o) \}$ since the second term in the rhs vanishes identically under reduction. Thus we can calculate $\{g_{ab}, t_{\hat\phi}S(N) \}$ and then, after the calculations are performed, input the substitution $\phi\mapsto \phi_o$ and $N\mapsto N_o$ arising from reduction. 

For the open manifold case, we want $\{g_{ab}, t_{\phi_o}B(N_o) \}$, and the argument relies on a slightly different property. Namely, that the bracket $\Delta=\{t_\phi(S(N)+B(N)),\pi_\phi\}$ possesses an inverse only for functions with compact support. Thus 
$\Delta(x)\Delta^{-1}(y)=\delta(x,y)$ only for $x,y\notin \partial\Sigma$.  Hence 
\be\label{equ:eom_open}
\{g_{ab}, t_\phi(S(N)+B(N)) \}_{\mbox{\tiny{DB}}}=\{g_{ab}, t_\phi(S(N)+B(N)) \}-\{g_{ab}, t_\phi S(N) \}_{|\Sigma-\partial\Sigma}
\ee 
Again, since the second term on the rhs vanishes upon reduction, we can calculate the bracket $\{g_{ab}, t_\phi(S(N)+B(N)) \}$ and then reduce $\phi\mapsto \phi_o$ and $N\mapsto N_o$.

\subsubsection{The equations of motion for maximal slicing}
Let us then write out the equations of motion of the Linking Theory for the open manifold case, and also the lapse fixing equation for the required slicing. For the open case, where we institute maximal slicing as opposed to constant mean curvature slicing (CMC), the canonical transformation of the metric variables are given by $(g_{ab},\pi^{ab})\mapsto (e^{4\phi}g_{ab},e^{-4\phi}\pi^{ab})$. What makes the equations of motion so tractable is that the conformal factor $\phi$ does not depend on the metric, as it does in the CMC case, where it is required to be total volume preserving. 

Using the canonical property of the transformation fairly simplifies the calculation of the equations of motion: 
$$\dot\pi^{ab}=\{\pi^{ab}, t_\phi\mathcal{H}(N, \xi^i)\}= e^{4\phi}\{t_\phi\pi^{ab}, t_\phi\mathcal{H}(N, \xi^i)\}=e^{4
\phi}t_\phi \{\pi^{ab}, \mathcal{H}(N, \xi^i)\}
$$
which gives: 
\begin{eqnarray}
\label{equ:eom_g}\dot g_{ab}&=&4\rho g_{ab}+2e^{-6\phi}\frac{N}{\sqrt g}(\pi_{ab}-\frac{1}{2}\pi g_{ab})+\mathcal{L}_{\xi}g_{ab}\\
\dot\pi^{ab}&=& Ne^{2\phi}\sqrt{g}\left(R^{ab}-2\phi^{;ab}+4\phi^{,a}\phi^{,b}-\frac{1}{2}R g^{ab}+2 \nabla^2\phi g^{ab}\right)\nonumber\\
&~&-\frac{N}{\sqrt g}e^{-6\phi}\left(2(\pi^{ac}\pi^b_c-\pi\pi^{ab})-\frac{1}{2}(\pi^{cd}\pi_{cd}-\frac{1}{2}\pi^2)g^{ab}\right)\nonumber\\
&~&-e^{2\phi}\sqrt {g}\left(N^{;ab}-4\phi^{(,a}N^{,b)}-\nabla^2N g^{ab}\right)+\mathcal{L}_{\xi}\pi^{ab}-4\rho\pi^{ab}\label{equ:eom_pi}
\end{eqnarray}
For the metric variables, upon reduction the Dirac bracket reproduces the Poisson bracket,\footnote{Note that this is not true for the variables that are being eliminated, namely $\phi$ and $\pi_\phi$. } and thus we get the possible equations of motion for Shape Dynamics simply by setting weakly $\pi\simeq 0$ and strongly $\phi=\phi_o$, $N=N_o$.

The lapse fixing equation  in the Linking Theory for open manifolds, which plays an important role in the reduction process,  is given  by the propagation condition on the lapse:
\be\label{equ:tphi_LFE_MS}\{t_\phi S(N),\pi_\phi\}=e^{-4\phi}(\nabla^2 N+2g^{ab}\phi_{,a}N_{,b}) -Ne^{-12\phi} \frac{G_{abcd}\pi^{ab}\pi^{cd}}{g}=0
\ee
where $t_\phi S(x)$ is given in \eqref{equ:LY}. 

There are many differences between the equations of Shape Dynamics, and its ``sister case", the conformal decompositions of ADM in  CMC slicings, which can be found for example in \cite{3+1 book}, in chapter 7. A first is that there the equations must include a reference metric, and the result is a preferred physical scale for each metric, unlike our own Weyl invariance 
\footnote{Furthermore, in Shape Dynamics the constraints decouple the conformal factor from the diffeomorphism constraint \cite{SD:FAQ}. This occurs because the diffeomorphism generator  decouples into one part that is the original diffeomorphism generator and another that gets absorbed by the conformal constraint upon reduction:
$$ \int d^3 x \pi^{ab}e^{-4\phi}\mathcal{L}_\xi (e^4\phi)g_{ab}=\int d^3 x \left(\pi^{ab}\mathcal{L}_\xi g_{ab}+4\pi\mathcal{L}_\xi \phi \right)
.$$}. A second is that there is no dilatation transformation elements, of the form $\rho\pi^{ab}$ and $\rho g_{ab}$. 

The remaining equations of motion for Shape Dynamics  for maximal slicing, besides \eqref{equ:eom_g} and \eqref{equ:eom_pi} with $\phi_o$ and $N_o$ solutions  substituted in, are just $\pi^{ab}_{;a}=0$ and $\pi=0$. We also pause to mention that equations of motion for Shape Dynamics in the CMC case are considerably more complicated, by the fact that the conformal factor $\hat\phi$ depends on $g$, and thus enters the variations. We will refrain from writing them down in the present paper, since we are not here concerned with the CMC case, but it is worth mentioning that the equations of motion for Shape Dynamics in CMC differ vastly from the equations of motion for conformal ADM in CMC, much more visibly so than when the two theories are in maximal slicing.  

\subsection{Restrictions on the Weyl invariance of Shape Dynamics}\label{sec:conformal_excision}

We now must pause for an intermezzo, in order to determine, in the case of manifolds with  boundary, exactly which Weyl transformations are still in the gauge group of Shape Dynamics. This will become important in obtaining the asymptotic charges.

For a closed spatial manifold  $\Sigma$, using the gauge-fixing of maximal slicing,  one has full dilatational (Weyl) symmetry at one's disposal (for phase space initial data with non-vanishing momenta). In many circumstances, however, we need to restrict the conformal factors we use to match the subset  explored in gauge-fixing the constraints. This intermezzo is aimed to show that there are subtleties in ascertaining which Weyl transformations are actually left over in Shape Dynamics. These subtleties can occur when one is dealing with the open manifold case and its boundary conditions,  the closed manifolds case with degenerate initial data in which the momenta vanish, or either of these when we want to leave some of the scalar constraints unfixed.

As we  saw, for CMC slicing the restriction that the leftover Weyl gauge transformations should preserve the total volume arises as a consequence of leaving one of the scalar constraints unfixed. This restriction manifests itself also as a  restriction on the Lagrange multipliers of the generator of Weyl transformations left over in Shape Dynamics for CMC, and can alternatively be obtained  immediately from the co-kernel of the lapse fixing equation.  The restriction in that case is what we call an ``inhomogeneity restriction" - the Lagrange multipliers suffer a projection $\rho\mapsto \rho-\mean{\rho}$, such that any constant $\rho$ does not translate into a Weyl transformation.

 Just as the kernel indicates the left over Hamiltonian generator, the cokernel indicates a ``non-traded" conformal factor, which is thus \emph{not}  a symmetry present in the final theory.\footnote{One can indeed straightforwardly  calculate the co-kernel of the lapse fixing equation $\{\pi_\phi(\rho), N(x)\}$ to be given by homogeneous $\rho$. The fact that the homogeneous cokernel differs from the $N_o$ kernel arises because for CMC foliation,  the lapse fixing equation is not self-adjoint.  }  Any restriction on the space of conformal fields  $\phi$  implies a restriction in the symmetry trading mechanism, and thus on the invariance leftover in Shape Dynamics.  

To be clear, the gauge-fixing $\pi_\phi=0$ does \emph{not} gauge-fix the Lagrange multiplier $\rho$, which is the multiplier of the Linking Theory constraint $t_\phi\pi_\phi\simeq 0$, and is what determines the range of the Weyl gauge transformations left in Shape Dynamics. The  restriction on the range of $\rho$ sneaks in  through the canonical transformation $t\phi$ performed in extended phase space, and this is connected to any restriction on the space of fields $\phi$.

Let us discuss this in slightly more detail.  We demand   that the solution  to the constraint we are  gauge fixing, let's say $t_\phi S(x)=0$,   be unique in the field space of  $\phi$ we are exploring, $C[\phi]$. Demanding uniqueness might implement some restriction on $C[\phi]$. From there, such a restriction in $\phi$ will migrate to the canonical transformation generator \eqref{equ:can_gen}. This can be explained as follows. Our construction of the Linking theory requires us first to trivially embed the original system into an \emph{unconstrained} (i.e. without any fall off conditions on the extra variables $\phi, \pi_\phi$) extended phase space. This does not alter the ADM constraints, but requires the additional constraint $\pi_\phi=0$. The $t_\phi$ that actually comes into the transformation of constraints is a canonical transformation in extended phase space, and it is this canonical transformation that must therefore be restricted. In the CMC case, this is manifest in the usage of $\hat\phi$, which is a projection of $\phi$ onto volume-preserving functions. 
  The restriction on the canonical transformation is then transmitted to a restriction in $\rho$, because the form of the extra constraint $t_\phi\pi_\phi =0$ obtained in the Linking theory  is itself altered to $\pi_\phi-4(\pi-\mean{\pi}\sqrt g)$. That is (see details of this calculation on \cite{SD1}), 
  \be\label{equ:rho_restriction} t_\phi\pi_\phi(x)=\diby {F[g,\hat\phi]}{\phi(x)}=\int d^3 x' 4\pi(x')\diby {\hat\phi(x')}{\phi(x)}\ee
   and so a restriction $\hat\phi$ - the hat now denoting a general restriction on the conformal factor -  usually will appear as a different form of the constraint $t_\phi\pi_\phi=0$. 

For fixed boundary conditions on the canonical transformations, no symmetry trading occurs on the boundary, and the Lagrange multiplier $\rho$ has to be set to zero there. If we fix $\phi$ on the boundary, we have $\delta\hat\phi(x)=0$, for $x\in \partial M$. Thus
 we have the expected   $t_{\hat\phi}\pi_\phi(x)=\pi_\phi(x)-4\pi(x)$ for $x$ in the interior of the manifold. 

More generally, a restriction of ${\hat\phi}$ to decay as  ${r^n}$ or faster implies that $\rho$ has to decay by the same order. This can be seen by writing $\hat\phi$ and $\phi$ in a power series in $r$. While $\hat\phi$ will have zero coefficients for the terms of order higher than those stipulated, $\phi$ will be unrestricted. Let us sketch a proof for analytic $\hat\phi$. Suppose (for $r>1$)
$$\hat\phi= \sum_{-\infty}^{n}a_i r^i 
$$
thus 
$$\diby {\hat\phi}{\phi}= {\delta_{a^i}^{b^i}}_{|\{i\leq n\}}
$$
Of course a similar expansion of the Lagrange multiplier $\rho=\sum_{-\infty}^{\infty}\rho_i r^i$ yields upon contraction with the generator $\diby {\hat\phi}{\phi}$, the projection $\rho\mapsto\sum_{-\infty}^{n}\rho_i r^i$. Thus only the terms of $\rho$ that decay faster than  $\delta\ln {\Omega}=\delta\phi$  produce non-trivial Weyl  transformations. 


\section{Boundary terms in Shape Dynamics for asymptotically flat boundary conditions}\label{sec:Boundaries}
Finally, we will calculate the asymptotic charges for Shape Dynamics for the asymptotically flat boundary conditions.

But first, we will motivate the study of this section by some considerations on the possibility of new symmetries over  Minkowski  initial data when Shape Dynamics is involved. Since the determination of the symmetries depends on solutions to differential equations (equations \eqref{equ:LY}  and \eqref{equ:tphi_LFE_abstract}) on an open manifold, one cannot analyze these symmetries independently of the consideration of boundary conditions. Furthermore,  as  established in section \ref{sec:conformal_excision}, fall-off conditions for the solution of \eqref{equ:LY} imply restrictions on the leftover symmetries in Shape Dynamics. There is thus a delicate interplay between all of these questions, which we attempt to illustrate now.

\subsection{Symmetries of the solution}

Let us consider  the curve of degenerate phase space data $(g_{ab}(t), \pi^{ab}(t))=(\delta_{ab}, 0)$, which  is already in maximal slicing. We call it degenerate because its momenta is identically zero, spoiling many of the uniqueness properties of the usual Shape Dynamics construction.

Over the set of data $(g_{ab},\pi^{ab})=(\delta_{ab},0)$, equation \eqref{equ:tphi_LFE_MS} gives:
\be\label{equ:LFE_Mink}e^{-4\phi}(\partial^2 N+2\delta_{ab}\phi_{,a}N_{,b}) =0
\ee
and the Lichnerowicz-York  equation, given by \eqref{equ:LY}, is 
\be\label{equ:LY_Mink} -8\partial^2\Omega=0
\ee
 In rectilinear coordinates $\{x^a\}$, the solutions to \eqref{equ:LY_Mink} for boundary conditions $\Omega\rightarrow \order{r}$  are generated by the linearly independent basis: $\Omega_o\in \mbox{span}[\{1,x^a\}]$, i.e. $\Omega_o(t)=c(t)+b_a(t)x^a$, where we denote solutions by the subscript $o$.\footnote{Note that whenever we include an index between parentheses it should be taken not as tensorial index, but merely an index parametrizing a given finite set.}


The general  solutions to  equation \eqref{equ:LFE_Mink} for $\Omega_o=c+\alpha_ax^a$ are of a complicated (yet closed) form.  Nonetheless, we will investigate here only two specific choices:  $\Omega=c$ and $\Omega=bx^a$. When $\Omega=c$,  if the boundary conditions for the lapse are also of $\order{r}$,  the solutions are given by $N_o^{(i)}=\{1, x^a\}$. When the conformal factor  is given by $\Omega=bx^a$, the solutions  is given  by the linear span of $\{1,x^{b\neq a}, 1/x^a\}$.\footnote{We are here assuming that the solutions are captured by the method of separation of variables. }

 Upon reduction the equations \eqref{equ:eom_g} and \eqref{equ:eom_pi}
give:
\begin{eqnarray}
\dot g_{ab}&=&4\rho \delta_{ab}+2\xi_{(a,b)}\nonumber\\
\dot\pi^{ab}&=& -N_{(i)}e^{2\phi^{(j)}}\sqrt{g}\left(-2\phi^{(j)}_{,ab}-4\phi^{(j)}_{,a}\phi^{(j)}_{,b}+2 \partial^2\phi^{(j)} \delta_{ab}\right)\nonumber\\
&~&-e^{2\phi^{(j)}}\sqrt {g}\left(N^{(i)}_{,ab}-2\phi^{(j)}_{(,a}N^{(i)}_{,b)}-\partial^2N _{(i)}\delta_{ab}\right)
\end{eqnarray}
Now,  since we have chosen  static phase space data as our ansatz, we must have $\dot g_{ab}=\dot\pi^{ab}=0$ for this to be a solution to Shape Dynamics. A more interesting solution, which could possess conformal symmetry, might be to look at the solutions over the phase space curve $(\rho(t)\delta_{ab},0)$, but we will leave this for further work.

 For  $\Omega=1$ and any $N_o^{(i)}={1, x^a}$  we automatically get that the momenta are preserved. It turns out that this is also true if we take $\Omega=x^a$, and the following smearings generate symmetries of the data (i.e $\dot g_{ab}=\dot \pi^{ab}=0$):
\begin{eqnarray}
P_{(c)}=(\rho, \xi^a_{(c)})=(0,\delta^a_{(c)}) ~&,& ~R_{(bc)}=(\rho, \xi^a_{(bc)})=(0,2 x_{[(b)}\delta^a_{(c)]})\nonumber \\
D=(\rho, \xi^a)=(c,-cx^a)~&,& ~K_{(c)}=(\rho, \xi^a_{(c)})=(2x_c,x^dx_ d\delta_{(c)a}-2x_{(c)}x_a)\label{equ:conf_sym}
\end{eqnarray}
where the index $(c)$ parametrizes the generators (for example, for each coordinate $x^a$), and should not be summed over (it is a fixed index). 
 This is a 10 dimensional group.

 By using the Shape Dynamics  algebra \eqref{equ:SD_algebra},  we can identify that this is indeed the 3-dimensional conformal group. Dilatations are given by $(\rho, \xi^a)=(c,-cx^a)$, special conformal transformations  by $(\rho, \xi^a_{(c)})=(2x_c,x^dx_ d\delta_{(c)a}-2x_{(c)}x_a)$,  translations along the $c$ coordinate by $\xi^a=\delta^a_{(c)}$  and rotations around the $c$ axis by $ \xi_{(c)}^a=\epsilon_{ad(c)}x^d$, which is equal to rotation along the $bc$ plane with generator $ x_{[(b)}{\delta^a}_{(c)]}$.  The non-zero part of the algebra is
\begin{eqnarray*}
[D, K_{(a)}]=-K_{(a)}~&,& ~[D,P_{(a)}]=P_{(a)}\\
~[K_{(a)}, P_{(b)}]=2\delta_{(bc)}D-2R_{(bc)}~&,&~[K_{(a)}, R_{(bc)}]=\delta_{(ab)}K_{(c)}-\delta_{(ac)}K_{(b)}\\
~[P_{(a)}, R_{(bc)}]=\delta_{(ab)}P_{(c)}-\delta_{(ac)}P_{(b)}~&,&~[R_{(ab)}, R_{(cd)}]=\delta_{(bc)}R_{(ad)}+\delta_{(ad)}R_{(cb)}-\delta_{(ac)}R_{(bd)}-\delta_{(dc)}R_{(ab)}
\end{eqnarray*}

 Furthermore, it can also  be checked that the algebra \eqref{equ:SD_algebra}  (for the smearings with $\rho=0$), we  reproduce exactly  the Poincar\'e algebra with boosts being given by $N^{(i)}=B_{(a)}=x^a$ and time translations by $N^{(i)}=T=1$ (the translations and rotations are left unchanged). In fact, there is a deeper story here, which we haven't touched upon yet. This is the occurrence of the central terms. In appendix \ref{appendix:central} we define these terms and show that they don't contribute.

\subsubsection{Inconsistent symmetries}

 The previous analysis would naively allow us to conclude that Shape Dynamics possesses an extended group of symmetries, composed of both 3-dimensional conformal symmetries and boosts and time translations. But upon calculating the algebra using \eqref{equ:SD_algebra}, we see that it does not close. More specifically, the brackets between the boosts and the special conformal transformations $[(x^a,0,0),(0,2x_c,x^dx_ d\delta_{(c)a}-2x_{(c)}x_a)]$ (where the order is given by $(N,\rho, \xi^a)$ ) fail to close, and do not constitute a symmetry of the system.

There are three different obstacles one needs to overcome in order for these two sets of symmetries to be compatible, and they all arise because we have to be more careful with the boundary conditions we choose for $\Omega$.   According to section \ref{sec:conformal_excision}, to allow transformations of the form $\rho=x^a$, the  phase space of $\Omega$ must be unrestricted to the order $\order{e^r}$. Furthermore it must have boundary conditions that grow even faster and yet determine the solution for $\partial^2\Omega=0$ uniquely.  If this can be done,  the unique solution to the Lichnerowicz-York equation  \eqref{equ:LY_Mink} with these asymptotics must still allow a solution to the lapse fixing equation  $N_o=x^a$,  with boundary conditions of the order $\order{r}$. The second issue is that the boundary conditions on the metric variables have to be invariant under whichever group of asymptotic symmetries we want to impose. However, asymptotic Minkowski boundary conditions are clearly not invariant under conformal symmetries. 
 The third and most important obstacle is  that if $\Omega$ is allowed to grow faster than $e^r$ it will be impossible to have finite counterterms that make the action differentiable, as we discuss in the next section. The conclusion is that to obtain true asymptotic conformal symmetry generators for Shape Dynamics we must consider a different set altogether of boundary conditions. We tackle this problem in an future publication.

\subsection{Asymptotically flat boundary conditions}

As we have mentioned more than once, the  Shape Dynamics Hamiltonian is a non-local implicit function. But Shape Dynamics is  obtained from a Linking theory, where all quantities are local. Thus the simplest way to  formulate Shape Dynamics boundary charges, counter-terms and fall-off conditions is to consider these in the larger setting of the Linking theory, and then use phase space reduction. The addition of extra variables is interesting in that it allows us to have more freedom in how we choose the boundary conditions. As this is but the first incursion in this rich landscape, we will confine ourselves to one consistent choice of boundary conditions. 

The total boundary variation can be calculated straightforwardly from the Linking theory:
\begin{multline}\label{equ:boundary_variations}\delta t_\phi B(N,\rho,\xi)=2\int_{\partial\Sigma} d^2 y\Big( \xi^ar^b\left(\pi^{cd}\left((5g_{ac}g_{bd}-g_{ab}g_{cd})\delta\phi+(g_{bd}\delta g_{ca}-\frac{1}{2}g_{ab}\delta g_{cd})\right)+g_{ac}g_{bd}\delta\pi^{cd}\right)\\
+\int_{\partial\Sigma} d^2 y\sqrt h e^{2\phi}\Big(8\left( N\delta\phi^{,a}-N^{,a}\delta\phi\right)r_ a+ \left( N\delta g_{ab;d}+(6\phi_{,d} N-N^{,d})\delta g_{ab}\right)(g^{de}g^{ab}-g^{da}g^{be}) r_e \Big)  
\end{multline}
Given fall-off conditions for all of our variables, some of the terms in \eqref{equ:boundary_variations} might vanish. The purpose of adding boundary terms is that their variation must cancel the remaining terms of \eqref{equ:boundary_variations}. The subtlety involved in the process is that the boundary terms themselves must be finite. It is this finiteness that  forbids us to have both 3-dimensional conformal symmetry and Poincar\'e symmetry simultaneously.

It is already possible to see that there will be some sort of mutual exclusion for  the fall-off conditions on the lapse $N$ and $\phi$ that allow both conformal and Poincar\'e symmetry.  Notice the term
$$ \int_{\partial\Sigma} d^2 y\sqrt h e^{2\phi}\Big(8\left( N\delta\phi_{,a}-N_{,a}\delta\phi\right)\Big)r^a
$$ 
As we know from section \eqref{sec:conformal_excision}, a fall-off rate for $\phi$  implies the same fall-off rate for the allowable Weyl gauge  transformations under which Shape Dynamics is invariant. Thus, to have spatial conformal symmetry, as in \eqref{equ:conf_sym}, we would need to allow $\phi\sim \order{r}'$, or $\phi \rightarrow \alpha_ a x^a+ a+ \order{r^{-k}}$, and thus $e^\phi\sim e^{r}$! Furthermore for the boosts we need a lapse to fall-off as $N\rightarrow \beta_a x^a +b+\order{r^{-k}}$. But a counterterm of the form:
\be   \int_{\partial\Sigma} d^2 y\sqrt h e^{2\phi}\Big(8\left( N\phi_{,a}-N_{,a}\phi\right)\Big)r^a
\ee imposes severe restrictions on the relative fall-off rate of $N$ and $\phi$. Even if we forget about the $e^{2\phi}$ term, and demand that the boundary conditions be such that $e^{4\phi}g_{ab}\rightarrow \delta_{ab}+\order{r^{-1}}'$ - which might be a natural  boundary condition defining waveless approximations -  finiteness  still demands that  the integrand $(N\phi_{,a}-N_{,a}\phi)\sim \order{r^{-2}}$ to maintain finiteness, and this cannot be obtained  if we allow both boosts and special conformal transformations.

\subsubsection*{Fall-off in extended phase space}

The usual boundary conditions taken for asymptotically flat ADM are \cite{RT, BO}:
\begin{eqnarray}
g_{ab}\rightarrow \delta_{ab}+\order{r^{-1}}''~~&,& ~~\pi^{ab}\rightarrow \order{r^{-2}}'\label{equ:fall-off}\\
N\rightarrow \order{1}'~~&,& ~~\xi^a\rightarrow \order{1}'\nonumber
\end{eqnarray}
where we designate $\order{r^{n}}'', \order{r^{n}}'$ as functions which fall off freely as $r^{n-1}$, but have possibly a contribution falling of as an even (odd) -parity  function in  $r^n$.  We say a function has odd parity if upon inversion $\mathbf{x}\rightarrow -\mathbf{x}$ it changes sign, and even parity if it doesn't. For example if $f\sim \order{r^{n}}'$ then 
$$f=h\left(\frac{\mathbf{x}}{r}\right)\frac{1}{r^n}+\order{r^{n-1}}
$$
such that $h\left(\frac{-\mathbf{x}}{r^n}\right)=-h\left(\frac{\mathbf{x}}{r^n}\right) $, and analogously for even parity.Any derivative will  reverse parity and increase the fall-off rate by one power in $r$, thus $\partial \order{r^{-k}}'= \order{r^{-k-1}}''$ and so on.  

The conditions on the lapse and shift above are called ``pure gauge", since for the appropriate counterterms the charges still vanish.  We will skip the derivation of the appropriate counterterms for the pure gauge conditions, and proceed directly to computation of the less restrictive boundary conditions which allow the Poincar\'e symmetries to emerge.

 For that we must loosen the fall-off conditions for the lapse and shift:
\be\label{equ:Lagrange_asympt} N\rightarrow  \alpha_{(a)}x^{(a)}+ c+ \order{1}'~~,~ ~~\xi^a\rightarrow \beta^{(c)}\delta^a_{(c)}+\mu_{(b)(c)}x^{[(b)}\delta^{(c)]a}+ \order{1}'
\ee
where again we have used indices under individual parentheses to  denote fixed indices. For Shape Dynamics, as we have seen, the lapse must asymptotically allow for the solutions of the lapse-fixing equation \eqref{equ:tphi_LFE_MS}. Interestingly, the solutions above are exactly those that fall-off to the orders $1$ and $r$. It could have occurred that Shape Dynamics disallowed the occurrence of charges for the boosts, but as with Minkowski, this is not what we happens.

 The fall-off condition for all the Lagrange multipliers, which are the smearings that define the foliation and conformal frame in the linking theory, must be such that the asymptotic conditions for the variables, \eqref{equ:fall-off}, are preserved through the evolution  equations \eqref{equ:eom_pi} and \eqref{equ:eom_g}. For the preservation of the asymptotic form of $g_{ab}$, this entails that $\rho\rightarrow \order{r^{-1}}''$, which also suggests the same fall-off for $\phi$, and thus a fall-off  $\Omega\rightarrow 1+ \order{r^{-1}}''$. 
\footnote{This brings us to the problem we have mentioned earlier: the boundary conditions for the metric variables are tailored so that they  still allow a Schwarzschild solution in phase space without allowing asymptotic wave propagation. These physical conditions should be translated into  Shape Dynamics, where the conformal mode of the metric and momenta is pure gauge. It seems that the right asymptotics for the metric would thus be that it approximates a conformally flat spacetime with traceless momenta. These spacetimes have been studied dynamically, and are also called Isenberg-Wilson-Mathews waveless approximation \cite{Isenberg}.  We will study these conditions in future work.  For now, since our main aim is to study the asymptotic Poincar\'e charges emerging in Shape Dynamics,  we stick with the usual asymptotic flatness conditions on the metric variables .} 

That the conditions \eqref{equ:Lagrange_asympt} preserve  \eqref{equ:fall-off} for the terms in \eqref{equ:eom_g} and \eqref{equ:eom_pi} that don't depend in $\phi$ is clear from the usual computations \cite{RT, BO}. Of these, the only non-trivial check is for the leading terms in $\mathcal{L}_\xi g_{ab}$, which is just $\xi_{(a,b)}$ and thus vanishes for the leading terms in \eqref{equ:Lagrange_asympt}, and thus falls off as $\order{r^{-1}}$. For the terms that \emph{do} depend on $\phi$, the terms $N\phi^{;ab}, ~\phi^{;a}N^{;b}$ and $ N^{;ab}$ all fall off as $\order{r^{-2}}$ and the remaining terms are easily seen to fall-off appropriately as well. Having verified that our boundary conditions are preserved by the equations of motion, we must now proceed to calculating the necessary boundary counterterms.

\subsection{Boundary Charges}
Now we are in a position to consider the fall-off rate of the different terms in \eqref{equ:boundary_variations}. Note that the normal $r^a$ has odd parity, as have the coordinate functions $x^a$.  For the reader's convenience, since we will be analyzing the different terms in \eqref{equ:boundary_variations}, we reproduce that equation here:
\begin{multline}\delta t_\phi B(N,\rho,\xi)=2\int_{\partial\Sigma} d^2 y\Big( \xi^ar^b\left(\pi^{cd}\left((5g_{ac}g_{bd}-g_{ab}g_{cd})\delta\phi+(g_{bd}\delta g_{ca}-\frac{1}{2}g_{ab}\delta g_{cd})\right)+g_{ac}g_{bd}\delta\pi^{cd}\right)\\
+\int_{\partial\Sigma} d^2 y\sqrt h e^{2\phi}\Big(8\left( N\delta\phi^{,a}-N^{,a}\delta\phi\right)r_ a+ \left( N\delta g_{ab;d}+(6\phi_{,d} N-N^{,d})\delta g_{ab}\right)(g^{de}g^{ab}-g^{da}g^{be}) r_e \Big)  
\end{multline}

 Let us consider first the terms that depend explicitly on $\phi$. Since $e^{2\phi}$  to first order is just unity, we can check that these are the usual boundary terms that one gets from the ADM boundary terms. Since the counter-terms for these are well-known, we simply refer to the literature \cite{RT, BO}. Here we will analyze the terms:
\begin{multline}\label{equ:phi_cterms} B(\delta\phi):=
2\int_{\partial\Sigma} d^2 y\Big( \xi^ar^b(\pi^{cd}((5g_{ac}g_{bd}-g_{ab}g_{cd})\delta\phi)\\
+\int_{\partial\Sigma} d^2 y\sqrt h e^{2\phi}r_e\Big(8( N\delta\phi^{,e}-N^{,e}\delta\phi)+ (6\phi_{,d} N\delta g_{ab})(g^{de}g^{ab}-g^{da}g^{be})  \Big)
\end{multline}
 
\begin{itemize}
\item For the first term: 
$$2\int_{\partial\Sigma} d^2 y\xi^ar^b(\pi^{cd}((5g_{ac}g_{bd}-g_{ab}g_{cd})\delta\phi)
$$  involves only the integral of  $r^c\pi^{ab}\delta\phi\sim \order{r^{-3}}''$, multiplied by $\xi^a$,  which thus vanishes asymptotically for $\xi^a=\delta_{(c)}^a$. For $x^{[(b)}\delta^{(c)]a}$ the integrand is of order $\order{r^{-2}}'$, and thus, being odd, vanishes when integrated over. Thus no counter-term is necessary to regularize this term. 
\item The second term
$$ 8\int_{\partial\Sigma} d^2 y\sqrt h e^{2\phi}r_e  N\delta\phi^{,e}
$$
 involves the integral of $r_e\delta\phi^{,e}\sim \order{r^{-2}}''$ (since $\delta\phi_{,e}\sim \order{r^{-2}}'$), which thus has odd parity for $N\sim x^a$ and vanishes. For $N\sim c$, a term of $\order{r^{-2}}''$ appears and thus requires a counter-term. There are a variety of counter-terms that would produce the same asymptotic charge, as for example
\be\label{equ:B_1} B_1:= 8\int_{\partial\Sigma} d^2 y\sqrt h r_e N\phi^{,e}
\ee
Note that this vanishes for $N\sim x^a$, since it is still of odd parity, and only produces a non-zero charge for $N=c$, a constant. One could instead  replace $\phi$ by $\Omega$ in \eqref{equ:B_1}, since $\Omega=e^\phi\sim 1+\phi$. Thus we could have 
\be\label{equ:B} B:= 8\int_{\partial\Sigma} d^2 y\sqrt h r_e N\Omega^{,e}
\ee

Another form of the asymptotic charge is given by  reinstating the $e^{2\phi}$ term completely:
\be\label{equ:B_2} B_2:= 8\int_{\partial\Sigma} d^2 y\sqrt h r_e N \Omega \Omega^{,e}
\ee
Here note that the leading contribution for $N r^e\delta\Omega \Omega_{,e}\sim \order{r^{-2}}'$ for $N\sim x^a$ and thus vanishes by parity. We will take the boundary charge then to be of the form given in \eqref{equ:B}
noting that the asymptotic  numerical value of \eqref{equ:B_1}, \eqref{equ:B_2} and \eqref{equ:B} are the same for fields obeying our boundary conditions. 

\item  For the third term
$$ -8\int_{\partial\Sigma} d^2 y\sqrt h e^{2\phi}r_eN^{,e}\delta\phi
$$
 we have an integrand of $r_e\delta\phi\sim \order{r^{-1}}'$ multiplied by $N^{,e}$, which has even parity for $N\sim x^a$, and thus the whole integral has odd parity and vanishes. For $N\sim c$, we have $N^{,e}\sim\order{r^{-1}}''$ which is still of even parity and thus the integral again vanishes. No counter-terms are needed for this. 
\item 
For the fourth and last term
$$ \int_{\partial\Sigma} d^2 y\sqrt h e^{2\phi}6\phi_{,d} N \delta g_{ab}(g^{de}g^{ab}-g^{da}g^{be}) r_e
$$
 we have an integrand of $r_e\phi_{,c}\delta g_{ab}\sim \order{r^{-3}}''$, which when multiplied by $N\sim x^a$ vanishes from parity, and for $N=c$ vanishes because it falls off as $r^{-3}$. Thus it requires no counter-term. 
\end{itemize}
We  that indeed we acquire a new asymptotic boundary term,  \eqref{equ:B}, which makes the variation well-defined and with finite charges. Note that Shape Dynamics does not interfere in the present case with the definition of either angular momenta or linear momenta;  only the energy  is affected.

 If we assume $N=c$ is a constant, we can write \eqref{equ:B}  as a volume term, which upon phase space reduction turns into
\be\label{equ:B_vol} {B}_{|N=c}= 8c\int_{\Sigma} d^3 x\sqrt g\nabla^2\Omega_o= c\int_{\Sigma} d^3 x\sqrt g\Omega_o(\left(R+\Omega_o^4\frac{\pi_{ab}\pi^{ab}}{g}\right))\ee

The total boundary is then given by the usual ADM boundary terms and our own \eqref{equ:B}:
\be\label{equ:SD_charges} M(N)+J(\xi)=\int_{\partial\Sigma}r_e\Big( 8N \Omega^{,e}+2\xi_ b\pi^{eb}+\sqrt h \left( N g_{ab,d}-N^{,d} g_{ab}\right)(g^{de}g^{ab}-g^{da}g^{be})\Big)
\ee
From \eqref{equ:SD_algebra}, and the fact that the propagating lapses are given exactly by $N_o=\{1, x^a\}$, from  \eqref{equ:SD_algebra} we precisely recover the Poincar\'e algebra. In fact, the story is slightly more complicated than this, as we might have obtained so called ``central terms" in the algebra. That this does not occur for our boundary conditions is shown in the appendix \ref{appendix:central}. 

\subsubsection{Shape Dynamics Energy} 

Although we have stipulated restrictive boundary conditions on the conformal factor and on the metric, we will now see that the notion of energy that arises from Shape Dynamics still possesses Weyl invariance. This invariance is not present in the ADM energy. 

For this proof, let us take  the usual ADM mass term and our term $B_1$ given in \eqref{equ:B_1} as the only terms  contributing to the energy.
We note that for the ADM mass, the only relevant term for the boundary charges is  \be\label{equ:boundary_ADM}-2\int_{\partial\Sigma}d^2 y\sqrt{h} (k-k_o)=\int_{\partial\Sigma}d^2 y \sqrt{h} (g_{ab,d}(g^{de}g^{ab}-g^{da}g^{be}))r_e
\ee
 where $k=h^{ab}r_{a;b}$ is the trace of the mean extrinsic curvature with respect to $g_{ab}$ of the boundary $\partial\Sigma$ and $k_o$ is not dynamical.  Thus the equation for the energy for an asymptotically flat solution of Shape Dynamics is given by:
\be\label{equ:SD_energy}
E_{\mbox{\tiny{SD}}}[g,\pi]=-\int_{\partial\Sigma}d^2y\sqrt{h}(2 (k-k_o)+8r_e \phi^{,e})
\ee 

Let us  calculate $t_\phi k$. The transformation of the Christoffel symbols is given by: 
\be\label{equ:christ_conf}
t_\psi\Gamma^e_{ab}=\Gamma^e_{ab}+2(\delta^e_ b\psi_{,a}+\delta^e_a\psi_{,b}-g_{ab}g^{ec}\psi_{,c})
\ee
If we consider the transformation of the normal to be given by $r^c\rightarrow e^{-2\psi}r^c$,
\be\label{equ:t_phi_k}2t_\psi k=2e^{-2\psi}(k+4(g^{ab}-h^{ab})r_a\psi_{,b}-2h^{ab}r_a\psi_{,b})=e^{-2\psi}(2k+8r^b\psi_{,b})
\ee 
 
Furthermore, we know that  the $\phi_o$ solution to equation \eqref{equ:LY} has the property that $\phi_o[e^{4\psi} g,e^{-4\phi}\pi]=\phi_o[g,\pi]-\psi$, since  equation \eqref{equ:LY} is itself obtained  by a Weyl transformation of the scalar constraint. Thus, if $\psi$ is a general conformal factor that obeys the boundary conditions imposed on $\phi$,\footnote{In fact, for the form of boundary charge \eqref{equ:B}, we get Weyl invariance even if we don't assume anything about $\psi$. } we have that 
\be 
E_{\mbox{\tiny{SD}}}[e^{4\psi}g,e^{-4\psi}\pi]=-\int_{\partial\Sigma}d^2y\sqrt{h}(2 (k-k_o)+8r_e (\psi^{,e}+(\phi^{,e}-\psi^{,e}))=E_{\mbox{\tiny{SD}}}[g,\pi]
\ee
which  shows that  \emph{the Shape Dynamics energy is fully Weyl invariant}.

\subsection{Shape Dynamics Hamiltonian in the asymptotically flat case }

As we mentioned in section \ref{sec:ps_red}, the evolution Hamiltonian for Shape Dynamics in maximal slicing is defined \emph{only} as a boundary contribution. Unlike what happens in ADM, the ``non-gauge", or evolution Hamiltonian is defined on the boundary for the entire phase space. On-shell, the Shape Dynamics Hamiltonian defines the energy of particular solutions, which is how we could have read \eqref{equ:SD_energy}. But this equation is valid  generally without the need of gauge fixation, so we define the Shape Dynamics Hamiltonian for maximal slicing in the asymptotically flat case as 
\be\label{equ:SD_H}
H_{\mbox{\tiny{SD}}}[g,\pi]=-\int_{\partial\Sigma}d^2y \sqrt h(2 (k-k_o)+8r_e \Omega_o^{,e})
\ee 
even when we are not on-shell. \footnote{We should note that the ADM Hamiltonian can also be defined on the boundary, given a phase space reduction induced by a gauge-fixing. The difference, which is a general difference between the maximal slicing gauge-fixing version of ADM and Shape Dynamics, is that the variables used for Shape Dynamics are still the full geometric variables $g_{ab}, \pi_{ab}$, whereas for the gauge-fixed version they are not. }

For different boundary condition on phase space we will then possibly have different Shape Dynamics' Hamiltonians. This begs the question related to the one that got us started into defining boundary terms in the first place: can the Shape Dynamics Hamiltonian be holographic, differentiable, and define non-trivial equations of motion? We cannot at this point prove a general assertion irrespective of the boundary conditions, but for the boundary conditions studied in this paper, we will now show that the emerging evolution Hamiltonian, namely \eqref{equ:SD_H} is differentiable \emph{and} provides non-trivial equations of motion on the bulk.  The technical reasons for this is that we can use the Gauss law for the extra boundary term \eqref{equ:B}, and the LY equation \eqref{equ:LY} to equate the Laplacian of $\Omega_o$ to a non-trivial volume term. 

For this,we re-write the extra term \eqref{equ:B} as a volume integral. We could then perform a linearization of the solution $\Omega_o$ (or of $\phi_o$) of \eqref{equ:LY}, and reinsert it into \eqref{equ:SD_H}. Although this is possible, it is laborious, and there is an easier way, that allows us to see  straightforwardly differentiability around a flat solution. Since this case already presents the only terms that could cause obstructions to differentiablity, we will refrain from the full argument of the general case (if one calculates the full linearization and inputs the boundary conditions, these are anyhow the only terms that survive).

We first take the variation of \eqref{equ:LY} around any solution of the ADM constraints, for which $\Omega_o=1$:
\begin{eqnarray*}
 t_\phi S(x) &=&\diby{t_\phi S(x)}{{g_{ab}(y)}}\cdot\delta g_{ab}(y)+
\diby{t_\phi S(x)}{{\phi(y)}}\cdot\delta \phi(y)\\
&=&\sqrt g\left(\nabla^2(g^{ab}\delta g_{ab})-\delta g_{ab}^{;ab}+(R^{ab}-\frac{1}{2}Rg^{ab})\delta g_{ab}\right)+\frac{2}{\sqrt g}(\pi^{ac}\pi^b_c-\frac{1}{4}\pi^{cd}\pi_{cd}g^{ab})\delta g_{ab}\\
&-&(6\frac{\pi^{cd}\pi_{cd}}{\sqrt g}+2R)\delta\Omega+8\nabla^2\delta\Omega=0
\end{eqnarray*}
where here $\delta\phi=\diby{\phi_o}{g_{ab}}$.
Taking the variation to be around a flat, static background $(g_{ab},\pi^{ab})=(\delta_{ab}, 0)$ the surviving terms are only:
\be
\delta t_\phi S_{|(\delta_{ab},0)}=\sqrt g\left(\nabla^2(g^{ab}\delta g_{ab})-\delta g_{ab}^{;ab}+8\nabla^2\delta\Omega\right)=0
\ee
Thus replacing in \eqref{equ:B}:
\be \delta B= 8\int_{\Sigma} d^3 x\sqrt g\nabla^2\delta\Omega_o= \int_{\Sigma} d^3 x\sqrt g\left(\nabla^2(g^{ab}\delta g_{ab})-\delta g_{ab}^{;ab}\right)
\ee
But the boundary term for this volume integral is exactly the opposite of the variation of \eqref{equ:boundary_ADM}, which thus cancels with the variation of the first term in \eqref{equ:SD_H}. That we take the variation around flat initial data is enough for these purposes, since our boundary conditions are taken around flat initial data.

\subsection{A simple ansatz for a solution of Shape Dynamics}

As a slightly non-trivial application of the formalism, we  briefly check that indeed, we recover the Schwarzschild mass from a very simple ansatz. First, using the equations of motion of Shape Dynamics we must construct an appropriate solution. Let us assume that the curve in  phase space data is still $(\delta_{ab},0)$, but that now we have a mass contribution at the center of our Universe. This is taken as a ``mass defect" at $r=0$.  As in \cite{SD:Causal, SD:Causal2}, for the correct coupling we assume that only the metric and the momenta scale with the conformal factor. 

As shown below, we will recover a Schwarzschild black hole in isotropic coordinates. This should come as no surprise, because in the present case of conformally flat metrics one can use the flat metric $\delta_{ab}$ as a fixed section in the conformal bundle (which means that the section won't enter the equations of motion). Furthermore,  it turns out that reducing the equations of motion of ADM to the subspace of conformally flat metrics coincides with calculating the equations of motion directly from a theory which only has the conformally flat metrics in its phase space. This latter theory is known as IWM (Isenberg-Wilson- Mathews) \cite{Isenberg}, and is known to reproduce ADM for spherically symmetric spacetimes \cite{3+1 book}. Nonetheless, it is instructive to pursue the solution straight from the Shape Dynamics ansatz. 

 In radial coordinates the LY equation \eqref{equ:LY} with the defect 
 becomes
\be\label{equ:Omega} \partial^2\Omega= 2\pi m\delta(r)
\ee
where the $2\pi$ factor is exactly what appears from the scalar constraint ($2\pi$ comes from $16\pi/8$, where the $8$ is the factor for $\nabla^2\Omega$ in \eqref{equ:LY}). The solution for equation \eqref{equ:Omega} with our boundary conditions is given by:
\be \Omega=1+\frac{m}{2r}
\ee

Of course now we must still verify that the curve $(g_{ab}(t),\pi^{ab}(t))=(\delta_{ab}, 0)$ is a solution to Shape Dynamics, i.e. we must check that $N_o, \rho, \xi$ exist that satisfy \eqref{equ:tphi_LFE_MS} and  $\dot g_{ab}=\dot\pi^{ab}=0$, according to \eqref{equ:eom_g} and \eqref{equ:eom_pi}. This is no easy task, and the solution, if it exists,  doesn't necessarily coincide with Schwarzcshild.\footnote{Furhtermore, the reconstruction of the metric is not necessary from the point of view of Shape Dynamics. More formally, it should emerge from equations of motion for test fields, and not merely by multiplying the spatial metric by the solution of the LY equation. }

The general solution to \eqref{equ:tphi_LFE_MS} with our boundary conditions is  $N=1-\frac{2b}{m+2r}$, where $b$ is an integration constant. 
The issue is that using the equations of motion \eqref{equ:eom_g} and \eqref{equ:eom_pi} we don't necessarily get $\dot g_{ab}=\dot\pi^{ab}=0$ for any choice of $b$ in the lapse, and thus the curve $(g_{ab}(t),\pi^{ab}(t))=(\delta_{ab}, 0)$ might not be a solution for Shape Dynamics. Equation \eqref{equ:eom_g} is automatically satisfied if $\rho=0, \xi^a=0$. If we input the lapse $N=1-\frac{2b}{m+2r}$ and $\phi=\ln{(1+\frac{m}{2r})}$ into \eqref{equ:eom_pi} we obtain
\be\dot\pi^{ab}=(1+\frac{m}{2r})^2
\left(
\begin{array}{ccc}
 \frac{8 (-1+b) m}{r (m+2 r)^2} & 0 & 0 \\
 0 & -\frac{4 (-1+b) m r \text{sin}^2(\theta )}{(m+2 r)^2} & 0 \\
 0 & 0 & -\frac{4 (-1+b) m r}{(m+2 r)^2} \\
\end{array}
\right)
\ee
Thus, the \emph{only} solution is obtained by taking $b=1$. To rewrite this solution as an ADM solution (and thus as a space-time), we must set the gauge such that $\Omega[\tilde{g},\tilde{\pi}]=1$.

 With this gauge-fixing in place, rewriting  $1-\frac{2m}{m+2r}=\frac{1-\frac{m}{2r}}{1+\frac{m}{2r}}$,  the reconstructed space-time from this is:
\be \label{equ:SD_new}
ds^2= -\left(\frac{1-\frac{m}{2r}}{1+\frac{m}{2r}}\right)^2dt^2+ (1+\frac{m}{2r})^4\left(dr^2+r^2(d\theta^2+\sin^2(\theta)d\varphi^2)\right)
\ee
which is exactly Schwarzschild in isotropic coordinates. Note that the lapse is uniquely fixed to be that given by $1-\frac{2m}{m+2r}$, but the spatial metric still would have the full Weyl ambiguity. To reconstruct the space-time metric, we have to go to the conformal gauge for which $\Omega=1$, which means choosing the conformal representative $\tilde g_{ab}=(1+\frac{m}{2r})^4\delta_{ab}$.  Thus we have shown that from a very simple ansatz, namely, flat space with a point-like mass source, Shape Dynamics reproduces the Schwarzschild solution in the ``preferred" isotropic coordinates.

For General Relativity, this solution in isotropic coordinated breaks down at the horizon (in these coordinates the horizon is located at $r=m/2$), since the 4-metric becomes degenerate there. The difference from GR is that for Shape Dynamics the solution is valid beyond the horizon,  all the way to the singularity itself. That is because the lapse is not identically zero, so geometrodynamical time itself carries on. Since the change of coordinates from isotropic to Schwarzschild breaks down at the horizon, the two spacetimes are identical outside the horizon, but are nonetheless physically different, i.e. not globally isometric.

\subsubsection{Recovering the Schwarzschild mass as a Shape Dynamics charge }

Trivially, one obtains   $\partial_r\phi\rightarrow-\frac{m}{2r^2}$, and asymptotically $k\rightarrow \partial_j g_{ij}-\partial_i(\delta^{kl}g_{kl})$ which is equal to zero since $g_{ij}=\delta_{ij}$.\footnote{Note that for the Shape Dynamics solution we must not impose the gauge such that $\Omega=1$.} Thus, for \eqref{equ:B} we are left with (after restoration of the $\frac{1}{16\pi}$ factor):
\be E_{\mbox{\tiny{SD}}}=-\frac{1}{2\pi}\int_{\partial\Sigma}d^2y\partial_r\phi (r^2\sin({\theta})d\theta d\phi)=m
\ee
and  using our modified energy charge \eqref{equ:B} we nonetheless obtain the correct Schwarzschild mass from a very simple ansatz.

 \section{Conclusions}\label{sec:conclusions}

We list below the main results of this paper. The results can be naturally  organized into two broader areas of investigation:

\subsubsection*{Formulation of Shape Dynamics for manifolds with boundary}

 Unlike General Relativity, the definition of Shape Dynamics requires phase space reduction,
  which can produce a non-trivial Hamiltonian that exists solely at the boundary, and is in that sense holographic. In General Relativity, the Hamiltonian exists in the bulk, and on-shell we are left with a boundary term representing the energy of the space-time. In Shape Dynamics, the entire evolution Hamiltonian is defined at the boundary, even off-shell. The energy is still just the value of the Hamiltonian over a particular solution. 
 
  In the general case, irrespective of the boundary conditions, we can only obtain a non-trivial Hamiltonian if the boundary conditions on the lapse allow for a non-zero lapse propagating the maximal slicing at the boundary. Luckily, the usual asymptotic conditions on the lapse that allow the definition of asymptotic boosts and energy are precisely those that propagate maximal slicing. This is the technical reason we are able to reproduce the Poincar\'e algebra for the charges (see last item).

\begin{itemize}

\item In general, boundary conditions on the unphysical conformal factor $\Omega$ that solves the LY equation \eqref{equ:LY} implies that symmetry trading will be limited. That is we do not obtain the full set of Weyl invariance in Shape Dynamics. We derived an explicit relation between  boundary conditions on the conformal factor $\Omega$ of the LY equation \eqref{equ:LY} and restrictions on the Lagrange multiplier $\rho$. The condition is that  the left over  Weyl transformations under which the theory is invariant are constrained to fall-off of  as $\ln{\delta\Omega}$. With this restriction, phase space reduction still proceeds as normal, but this is one of the reasons we do not obtain charges for the Weyl transformations at infinity. 

\item For asymptotically flat conditions, we calculated the boundary terms  emerging from the requirement of differentiablity and finiteness of the charges in Shape Dynamics.  By a careful study of the boundary charges, we have derived an extra term for the boundary Shape Dynamics Hamiltonian (as compared to the ADM energy of an asymptotically flat space-time). This implies an extra term for the definition of the Shape Dynamics energy charge,  but we find  no extra contribution for the angular momenta or linear momenta. 

\item This extra term ensures that the evolution Hamiltonian (and thus also the energy) for Shape Dynamics is  Weyl invariant. 

\item  The algebra of the constraints emerging for these boundary conditions are calculated and shown not to produce central terms. We have found that for the boundary conditions studied here, the boundary charges still obey the Poincar\'e algebra.

\item For different boundary conditions on phase space we can obtain different Shape Dynamics' Hamiltonians. But can the (Shape Dynamics) Hamiltonian be differentiable, ``holographic", and provide non-trivial equations of motion? We cannot at this point prove a general assertion irrespective of the boundary conditions, but for the boundary conditions studied in this paper, by using a linearization of the LY conformal factor and the Gauss law we have shown that the emerging evolution Hamiltonian, namely \eqref{equ:SD_H} satisfies all these requirements.

\end{itemize}

\subsubsection*{Equations of motion}

As mentioned, upon the occurrence of boundaries, phase space reduction allows for a remaining Hamiltonian evolution generator at the boundary. The leftover Hamiltonian $\mathcal{H}$ that is left at the boundary is a functional of the implicit function $\phi_o[g,\pi]$. This makes the equations of motion derived from $\diby{\mathcal{H}}{g_{ab}}$ and $\diby{\mathcal{H}}{\pi^{ab}}$  impossible to carry out explicitly. This is where we can use phase space reduction to our advantage: we can obtain the equations of motion individually over each solution of $\phi_o$, and do not need to perform functional differentiation of an implicit function. 

\begin{itemize}

\item We presented the equations of motion for Shape Dynamics in the open, maximal slicing case. They  differ in some respect  from its closest cousin, conformal ADM in maximal slicing.\footnote{The equations of motion for the CMC case are vastly different than the ones for conformal ADM in CMC slicing, but since they are also much more laborious and we are not using them in this paper, we have refrained from writing them down here. } One of the main differences is that one does not in any way couple the diffeomorphism constraint to the conformal factor. This is related to the fact that Shape Dynamics does not need either an auxiliary background metric or the use of odd density weight in its formulation, because the conformal factor is not part of the metric, but an auxiliary field (see also \cite{SD:FAQ}).  This opens a new possibility in numerical relativity.


\item The equations of motion are used to show that over   Minkowski initial data, Shape Dynamics might be naively taken to  have both 3-dimensional conformal symmetry and Poincar\'e symmetry. This misconception is due to the fact that the propagating lapses for Shape Dynamics allow for exactly the right lapses to produce boosts and time translations. However, when we properly consider the preservation of the boundary conditions from the equations of motion and the finiteness of the boundary terms, we find these two distinct sets of symmetries mutually exclusive.

\item We formulated a very simple ansatz for Shape Dynamics,  consisting of flat, static space with a point-like mass at the center of the Universe. We then used the equations of motion of Shape Dynamics to find a solution to this  ansatz. The solution cannot be inferred merely from finding the unique solution to the Lichnerowicz-York conformal factor and one must use the full power of the equations of motion.  For our solution to this ansatz, the correction term to the energy provides the full mass, since the metric contribution is zero.  A simple reconstruction of space-time from our solution  finds exactly Schwarzschild in isotropic coordinates. Using the Shape Dynamics definition of energy we still obtain the Schwarzschild mass.

\end{itemize}

Although we have obtained a Weyl invariant mass charge, we have not been able to obtain here a phase space that allows asymptotic conformal symmetry generators (or charges), although they would seem to naively be also present in the Shape Dynamics solution over Minkowski initial data in phase space. We attribute this to three interconnected facts  (which we explain in the main text)  all arising from the same root: the definition of the asymptotic boundary values for the metric variables is poorly adapted to Shape Dynamics. Two further investigations are thus called for: i) the loosening of the boundary conditions to something akin to the IWM waveless approximation \cite{Isenberg}, and ii) the study of asymptotically AdS boundary conditions.

As a last remark, and a  matter of curiosity,   there is a sense in which spatial special conformal transformations are dual to boosts, and dilatations to time translations. The technical reason for this is that Poincar\'e symmetry is related to the kernel of the phase space reduction matrix (Dirac matrix), whereas 3-dimensional conformal symmetry is generated by elements coinciding with the co-kernel. This means roughly that you can only get one at the expense of the other.  
At least for vacuum  Shape Dynamics on closed spatial manifolds,  this fact would make these two sets of symmetries existing in the flat torus case formally  mutually exclusive: Poincar\'e invariance prohibits spatial conformal invariance, and vice-versa. That is, for odd-dimensional compact manifolds, the Atiyah-Singer index theorem \cite{Atiyah-Singer} demands that any differential operator, for instance $\{\pi_\phi(x),t_\phi S(y)\}$, have zero index, which means that the dimension of the kernel always equals the dimension of the co-kernel. Thus if one has a certain number of lapses propagating the CMC condition this entails that we must have the same number of Weyl transformations disallowed from the remaining Shape Dynamics invariance group.  This motivates us to conjecture that the two sets of symmetries, Poincar\'e and 3-dimensional conformal (i.e. dilatations and special conformal transformations), are \emph{always}  mutually exclusive.

\begin{appendix}

\section{The difference between the open and closed case in the treatment of the cokernel of the Lapse fixing equation}\label{appendix:LFE_CMC}
Now we expand on a different point of view on the construction of Shape Dynamics: the infinitesimal one.  

  The infinitesimal criteria for gauge fixing is transversality. I.e. to have one phase space function $\Phi$  gauge-fix another, $\Psi$,   the Poisson bracket between them should be weakly invertible (non-vanishing Fadeev-Popov determinant). In the case at hand, we can easily verify that for the gauge-fixing function $\pi_\phi=0$ and first class constraints of the Linking theory, the only weakly non-vanishing Poisson-bracket is
\be\label{equ:Pb_pi_phi,pi}\{t_\phi S(N),\pi_\phi(x)\}=4 t_\phi\{S(N),\pi(x)-\mean{\pi}\sqrt g(x)\}\ee
This means that, if there are $N$ such that 
\be\label{LFE}\{S(N),\pi(x)-\mean{\pi}\sqrt g(x)\}=0\ee
 the Poisson bracket is not invertible. In other words, the Fadeev-Popov determinant related to the gauge-fixing vanishes. Equation \eqref{LFE} will be henceforth called the Lapse Fixing Equation (LFE). If there is generically a unique solution to \eqref{LFE} one can show by going to the ``solution section" of the conformal factor that the kernel of $\{t_\phi S(N),\pi_\phi(x)\}$ is completely captured by the kernel of the LFE.  

 It turns out that we can rewrite \eqref{LFE} as 
\be\label{equ:Delta_constr}\{S(N),\pi(x)-\mean{\pi}\sqrt g(x)\}=\diby{t_\phi S}{\phi}_{|\phi=0}\cdot (N)= \Delta  N-\mean {\Delta  N}=0
\ee
where $\Delta$ is a differential operator functionally dependent on $(g,\pi)$, which is furthermore invertible when $\pi^{ab}\neq 0$. Thus there is such a non-trivial  solution $N_o[g,\pi]$. This means that  out of an infinite set of scalar constraints (one for each point), we are left with one that remains first class wrt our gauge-fixing.  With the lapse smearing $N=N_o$ ``excised" from the space of smearings, the Poisson bracket of the gauge-fixing with the remaining scalar constraints  form an operator that is now invertible (non-vanishing Fadeev-Popov determinant) and thus is a true gauge fixing of the remaining generators. 

So far, there is nothing really new in this analysis, but a few comments are in order. 
Before reduction, there is an equivalence: 
$$
\left(\pi_\phi-4(\pi-\langle\pi\rangle\sqrt g)\right)(\rho)= (\pi_\phi-4\pi)(\rho-\mean{\rho})
$$
So the only effectively non-trivial smearings we can choose, i.e. those that will have a non-trivial action on the canonical variables, are those that are not constant. This restriction corresponds to the fact that we are limiting ourselves, in the closed case, to Weyl transformations that keep the total volume fixed. Non-constant smearings here are the infinitesimal version of volume-preserving Weyl transformations.

Suppose however that we did not know beforehand which conformal factors were being excluded, but only had access to the bracket: 
 \be\label{equ:Delta_cokernel}\{S(N),(\pi(x)-\mean{\pi}\sqrt g)(\rho)\}= \mean{\rho\Delta  N}-\mean {\rho}\mean {\Delta  N}
\ee
Using integration by parts, one can check that the co-kernel is given simply by $\rho=\mean \rho$, which is consistent with our previous comments.\footnote{In this case  the  bracket is not a self adjoint operator between the domain of $\rho$ and $N$, which allows the kernel and co-kernel to be distinct, albeit with the same dimensionality. } 

The gauge-fixing  $\pi_\phi=0$ clearly \emph{does not} gauge fix the constraint $\left(\pi_\phi-4(\pi-\langle\pi\rangle\sqrt g)\right)(x)$, since their Poisson Bracket vanishes. Hence the Lagrange multiplier $\rho$ is left arbitrary after reduction. However, since we were forced to consider volume preserving conformal transformations to get uniqueness of $\phi$ from  the non-homogeneous version   \eqref{equ:non_homogeneous LY} of the LY equation \eqref{equ:LY}, the smearing  always appears as $\rho-\mean\rho$.

This is the situation for the construction over closed spaces:  to "shave off" redundancy in the solutions of  equation \eqref{equ:non_homogeneous LY} we change the range of conformal factors we allow (e.g. to be volume preserving), ultimately changing the form of the constraint itself (i.e. $4\pi\rightarrow 4(\pi-\mean{\pi})$. \emph{Limiting the space of conformal factors to lie in the  (integral manifold of the) complement of the cokernel of \eqref{LFE} is what allows us to obtain a unique solution to  \eqref{equ:non_homogeneous LY}}. 

   However, in the open case we can obtain a unique solution to the full LY equation \eqref{equ:LY}  \emph{and} retain a global Hamiltonian merely by choosing boundary conditions on our conformal factors, and thus do not need to change the form of the constraints. This fact is what will allow us to retain the full range of Weyl transformations even in the face of non-trivial co-kernels of the lapse fixing operator.

To summarize, in the usual construction of Shape Dynamics for closed manifolds, one chooses CMC as opposed to maximal slicing, which is meant to allow for non-trivial time evolution, i.e. it allows \emph{one} lapse to remain unfixed. In that case, it turns out that this is achieved by excluding one possible conformal transformation, the homogeneous one. In fact the CMC condition translates into the gauge symmetry only allowing non-homogeneous conformal factors, i.e. conformal factors of the form $\rho-\mean{\rho}$. 
On the other hand, still in the closed manifold case, one could also choose maximal slicing, and  thus remove the former restriction on the conformal factor.  This is in fact similar to what York et al do. By removing the restriction on the conformal factor, the one unfixed lapse generating time evolution is set to zero, and we are left only with an initial value formulation. In other words, by leaving $N_o$ unfixed (since it is in the kernel of the lapse fixing operator), we must not include the conformal transformations that are in the co-kernel of the lapse fixing operator.

Over a ``Minkowski curve" something very similar occurs. Again, naively the best-matching gauge-fixing $\pi_\phi=0$ might turn out to have a kernel $N_o^{(i)}$ and co-kernel $\rho_o^{(i)}$, where $i$ runs over a finite set. In the Minkowski case, the operator turns out to be self-adjoint, and thus we can identify $N_o^{(i)}$ with $\rho_o^{(i)}$.  But as we have explained there is a fundamental asymmetry in the phase space reduction process: the kernel of the operator signals which of the lapses \emph{remain} as generators of dynamics, whereas the cokernel signals which conformal smearings are \emph{disallowed} by the symmetry trading\footnote{We should mention here that the conformal factors that are leftover are also going to be responsible for true dynamics of the Universe, as in the formulation of the global Hamiltonian by a volume constraint \cite{SD:FAQ}.}

Hence, upon allowing  $N_o^{(i)}$ to remain as the generators of dynamics, one  must excise  $\rho_o^{(i)}$. Accidentally, these are \emph{exactly} the conformal factors that were needed to forge our 3-dimensional conformal symmetry! The conformal factor $\rho=c$ was necessary for dilatations and $\rho=x^a$  for special conformal transformations. Thus if we follow what occurs in the closed topology case, this would mean that for this particular solution we would have to excise exactly the conformal factors that we wanted. But it also suggests that we could augment our system so that we are left only with a trivial kernel of the gauge-fixing matrix, in which case, if the equations of motion were left unchanged, it would allow us to recover the 3-dimensional conformal symmetry in detriment of boosts and time translations.

\section{Auxiliary calculation of the boundary terms} \label{app:Boundaries} 

We start with writing out the respective scalar constraint in the Linking theory: 
\be\label{equ:LT_scalar}
t_\phi S(N)=\int d^3 x N\left( e^{-6\phi}\frac{G_{abcd}\pi^{ab}\pi^{cd}(x)}{\sqrt g}-e^{2\phi}\sqrt g(x)\left(R(x)-8(\phi^{,a}\phi_{,a}+\nabla^2\phi)\right)\right) 
\ee
Clearly for the $\delta_gt_\phi S$ boundary terms we will get all of the ones from the usual $R$ variation, but here substituting $N\rightarrow Ne^{2\phi}$:
$$\int_{\partial\Sigma}d^2 y \sqrt h \left(e^{2\phi}Nh^{ab}\left(\delta g_{ab;c}-\delta g_{ac;b}\right)-(e^{2\phi}N)_{;c} h^{ab}\delta g_{ab} +(e^{2\phi}N)^{;b} \delta g_{bc}  \right) r^c
$$
We also have the extra  first derivatives of the metric arising from the Christoffel symbols in $\nabla^2\phi$, namely $ -8e^{2\phi}N\phi_{,c}\Gamma^c_{ab}g^{ab}$. The boundary term arising from this is going to be given by
\begin{multline}-\int d^3 x \sqrt g 8e^{2\phi}N\phi^{,c}g^{ab}\left(\delta g_{bc;a}-\frac{1}{2}\delta g_{ab;c}\right) =
-\int_{\partial\Sigma} d^2 y \sqrt h 8e^{2\phi}N\phi^{,c}\left(\delta g_{bc}r^b-\frac{1}{2}h^{ab}\delta g_{ab}r_c\right)-\{\mbox{vol. term}\}
\end{multline}
The total boundary term  for the metric variation is then:
\be\label{equ:LT_scalar_boundary_g}
\delta_g t_\phi B(N)=\int_{\partial\Sigma}d^2 y \sqrt h e^{2\bar\phi} \left(\bar N\delta g_{ab;d}+(6\phi_{,d}\bar N-N^{,d})\delta g_{ab}\right)(g^{de}g^{ab}-g^{da}g^{be}) r_e   
\ee
where we have denoted with an over bar the choice of boundary functions for $\phi$ and $N$, and by $h_{ab}$ the boundary value of the metric, as usual. 

Now for the $\phi$ variation, we have the boundary terms coming from:
\begin{multline}
8\int d^3 x \sqrt g Ne^{2\phi}\left(2\phi^{,a}\delta\phi_{,a}+\nabla^2\delta\phi\right)=\\
-8\int d^3 x\sqrt g e^{2\phi}\left(2(N\nabla^2\phi+\phi_{,c}N^{,c}+2N\phi_{,c}\phi^{c})\delta\phi+(N^{,c}+2N\phi^{,c})\delta\phi_{,c} \right)
+8\int_{\partial\Sigma} d^2 y\sqrt h Ne^{2\phi}\left(2\phi^{,a}\delta\phi+\delta\phi^{,a}\right)r_ a\\
=8\int_{\partial\Sigma} d^2 y\sqrt h e^{2\phi}\left(N\delta\phi^{,a}-N^{,a}\delta\phi\right)r_ a+\{\mbox{volume terms}\}
\end{multline}
So the total boundary term for the conformal variation of the scalar constraint is then:
\be\label{equ:LT_scalar_boundary_phi}\delta_\phi t_\phi B(N)=8\int_{\partial\Sigma} d^2 y\sqrt h e^{2\bar\phi}\left(\bar N\delta\phi^{,a}-N^{,a}\delta\phi\right)r_ a
\ee 
where, to emphasize, $r^c$ is not the normal to the boundary according to $t_\phi g_{ab}$ but according to $g_{ab}$.

For the momentum constraint we have:
\be\label{LT_momentum}
t_\phi H_a(\xi^a)=\int d^3 x({\pi_a^b}_{;b}+10\pi_a^b\phi_{,b}-2\pi\phi_{a})\xi^a
\ee
The total boundary term for the momentum constraint is then:
\be\label{equ:LT_momentum_bdary}
\delta t_\phi B_a(\xi^a)=2\int_{\partial\Sigma} d^2 y \xi^ar^b\left(\pi^{cd}\left((5g_{ac}g_{bd}-g_{ab}g_{cd})\delta\phi+(g_{bd}\delta g_{ca}-\frac{1}{2}g_{ab}\delta g_{cd})\right)+g_{ac}g_{bd}\delta\pi^{cd}\right)
\ee

\section{Central terms in the algebra}\label{appendix:central}

Since integration by parts is used to obtain the form of the brackets given in \eqref{equ:ADM_algebra}, we get extra boundary terms that do not arise from variation of the fields on the boundary. Namely
\be\{\bar H^\mu(\xi_\mu),\bar H^\nu(\eta_\nu)\}=\{H^\mu(\xi_\mu),H^\nu(\eta_\nu)\}+\{B^\mu(\xi_\mu),B^\nu(\eta_\nu)\}
\ee
Although the variations are taken without boundary, we will still have a difference
\be\label{equ:Q}
Q(\eta^\nu, \xi^\mu):=\{H^\mu(\xi_\mu),H^\nu(\eta_\nu)\}-H([\xi, \eta]_{\mbox{\tiny HD}})\ee
 due to integration by parts.

 In \cite{Carlip}, it is shown that it is these terms that are relevant for the computation of the central term of the computation of the central term in a Virasoro algebra on the boundary. By lack of a better name, we will call these terms, $Q(\eta^\nu, \xi^\mu)$, \emph{the boundary terms of the deformation algebra}. The computation of $Q$ does not involve a specific form of the boundary terms, since we will be throughout assuming that there always exists a given extra boundary term can be added to cancel the boundary variations of the fields, whichever boundary conditions we choose. 

Let us briefly calculate the boundary deformation algebra for the usual ADM constraints. In the following section we show that indeed, by the canonical properties of our  transformation in the Linking theory,  the boundary deformation algebra for the Linking theory is just given by a conformal transformation of the boundary deformation algebra for the ADM system. This allows us to see straightforwardly that the relevant terms that might interfere with the algebra of \eqref{equ:SD_charges} vanish for our asymptotic boundary conditions. 

\subsection{$\{S,S\}$ term}

The only non-trivial Poisson bracket for ADM, since as we saw the momentum constraint generate only 3-diffeomorphisms, is $\{S(x),S(y)\}$. For the smeared version, all the terms that are both linear in the smearings will cancel out upon anti-symmetrization, so in the end we only have to calculate:
\be \int d^3x N_2\left(2g_{ec}g_{fd}\pi^{cd}-g_{ef}\pi\right)\left(-g^{ef}\nabla^2N_1+
N_1^{;ef} \right)=2\int d^3x N_2\pi^{cd}{N_1}_{;cd}
\ee
And thus
\begin{multline}\label{equ:SS_algebra}\{S(N_1),S(N_2)\}= \int d^3 x\pi^{cd}\left(N_1N_{2;cd}-N_2N_{1;cd}\right)\\=H^a(N_1\nabla_a N_2-N_2\nabla_a N_1)+\int_{\partial\Sigma}d^2y (\pi^{cd}(N_1N_{2,c}-N_2N_{1,c})r_d)
\end{multline}
where the anti-symmetrization cancelled the mixed derivatives, and $r_a$ is the normal to $\partial\Sigma$. 

The term we will use in the main text is the boundary of the Poisson bracket: 
\be\label{equ:SS_bdary}
Q(N_1,N_2):= \int_{\partial\Sigma}d^2 y\pi^{cd}(N_1N_{2,c}-N_2N_{1,c})
\ee
 This term is already contained in the boundary charges. 

\subsection{$\{H^a,H^a\}$ term}

Now that we have all the intermediary calculations out of the way, this bracket will be more straightforward. Before variations we are allowed to do integration by parts in whichever way we want, since we assume that there exist boundary terms whose variation will cancel the variations on the boundary. Thus we can write
\be \label{equ:HH}
\{H_a(\xi_1^a),H_b(\xi_2^b)\}=\left\{\int d^3x( \pi^{ab}\mathcal{L}_{\vec\xi_1} g_{ab}),\int d^3x'( \pi^{ab}\mathcal{L}_{\vec\xi_2} g_{ab})\right\}=\int_\Sigma (\mathcal{L}_{\vec\xi_1} g_{ab}\mathcal{L}_{\vec\xi_2}\pi^{ab}- 1\leftrightarrow 2)d^3x
\ee upon integration by parts we obtain: 
\be\label{equ:HH_algebra}\int_\Sigma \pi^{ab}(\mathcal{L}_{\vec\xi_1}\mathcal{L}_{\vec\xi_2}-\mathcal{L}_{\vec\xi_2}\mathcal{L}_{\vec\xi_1})g_{ab})+\int_{\partial\Sigma}(\mathcal{L}_{\vec\xi_1}g_{ab}\xi^c_2 \pi^{ab}r_c- 1\leftrightarrow 2)
\ee
Since $[\mathcal{L}_{\vec\xi_1},\mathcal{L}_{\vec\xi_2}]=\mathcal{L}_{[\vec\xi_ 1,\vec\xi_2]}$ we get the respective part of the boundary deformation algebra:
\be\label{equ:HH_bdary}Q(\vec\xi_1, \vec\xi_2)=\int_{\partial\Sigma}d^2 y\left(\mathcal{L}_{\vec\xi_1}(g_{ab})\xi^c_2 \pi^{ab}-\mathcal{L}_{\vec\xi_2}(g_{ab})\xi^c_1 \pi^{ab}+  [\vec\xi_ 1, \vec\xi_ 2]_a\pi^{ac}\right)r_c)
\ee
This term is also already contained in the boundary charges. 

\subsection{$\{H^a, S\}$ term}
As in the main section the gauge fixing will impose $\frac{\pi}{\sqrt g}=0$, and $\pi$ comes in squared in the scalar constraint, we can set it to zero even before variation, which is what we do here. No complications would arise had we left it in, and we do it to be more compact. 
\begin{multline} \{H_a(\xi^a),S(N)\}=\\
\int d^3x\left(2\frac{\pi_{ab}}{\sqrt g}\mathcal{L}_{\vec\xi}\pi^{ab}N+\mathcal{L}_{\vec\xi}g_{ab}(-\frac{1}{2}\frac{\pi^{cd}\pi_{cd}N}{\sqrt g}g^{ab}+2\frac{\pi^{ca}\pi_{c}^bN}{\sqrt g}+\sqrt g(-\frac{1}{2} RN g^{ab}+R^{ab}N+g^{ab}\nabla^2N-N^{;ab}))\right)
\end{multline}
The first three terms, involving the momenta, sum to
$\mathcal{L}_{\vec\xi}(\frac{\pi^{cd}g_{ce}g_{df}\pi^{ef}}{\sqrt g})N
$. After writing $\mathcal{L}_{\vec\xi}g_{ab}=2\xi_{(a;b)}$, doing integration by parts, and using the Bianchi identity: $R^{ab}_{;a}=\frac{1}{2}R^{,b} $, we get for the term
\begin{multline}\int_{\Sigma} d^3 x\sqrt g 2\xi_{(a;b)}(-\frac{1}{2} RN g^{ab}+R^{ab}N+g^{ab}\nabla^2N-N^{;ab})=\\
\int_{\Sigma} d^3 x\sqrt gR\xi^cN_{,c}+\int_{\partial\Sigma}d^2 y\sqrt h(-\frac{1}{2}RN\xi^a+\xi^cR_c^{\phantom{c}a}N+\xi^c_{\phantom{a};c} N^{;a}-\xi^a_{\phantom{a};c}N^{;c})
\end{multline}thus 
\be\label{equ:HS_algebra}
 \{H_a(\xi^a),S(N)\}=S(\mathcal{L}_{\vec\xi}N)+Q(\xi^a, N)
\ee
where 
\be\label{equ:HS_bdary}
Q(\xi^a, N)=\int_{\partial\Sigma}d^2 y\sqrt h(-\frac{1}{2}SN\xi^a+\xi^cR_c^{\phantom{c}a}N+\xi^c_{\phantom{a};c} N^{;a}-\xi^a_{\phantom{a};c}N^{;c})
\ee
is the respective part of the boundary deformation algebra.  

The only terms that do not obviously vanish in our asymptotic conditions, once we perform the canonical transformation, are $t_\phi (\xi^c_{\phantom{a};c} N^{;a}-\xi^a_{\phantom{a};c}N^{;c})$ . But $t_\phi(\xi^a_{\phantom{a};c})\sim\xi^a_{\phantom{a},c}+\xi^a\partial\phi$ and this vanishes as $\delta^a_c + \order{r^{-1}}''$. In the same way $N_{,c}$ vanishes as $\delta^a_b+\order{r^{-1}}''$. Thus when multiplied by $r_e$, which is odd, the total integrand is always  odd for the relevant powers of $r$ and thus vanishes. 

\subsection{Boundary deformation algebra for the Linking theory}

It is easy to check that our canonical transformation is indeed still canonical even in the presence of boundary terms. This means that in calculating the boundary terms, we can use the identity:
\be\label{equ:canonical_PB's}
 \{t_\phi f_1(g,\pi), t_\phi f_2(g,\pi)\}=t_\phi\{f_1(g,\pi),  f_2(g,\pi)\}
\ee
and in particular we get for the boundary of the deformation algebra:
$t_\phi Q(\xi_1^\mu, \xi_1^\mu)$. We explicitly check these identities for a few cases for illustration in the appendix. 

 The extra constraint obtained in the Linking Theory through the Stuckelberg  mechanism becomes $C(x)=\pi_\phi-4\pi\approx 0$. Since neither $\pi_\phi$ nor $\pi$ possess any spatial derivatives, regardless of the presence of boundaries a short computation shows that the following identity still holds: 
\begin{subequations}
\begin{eqnarray}
\{\pi_\phi, t_\phi g_{ab}\}&=& t_\phi\{4\pi, g_{ab}\}\\
\{\pi_\phi, t_\phi \pi^{ab}\}&=& t_\phi\{4\pi, \pi^{ab}\}
\end{eqnarray}
\end{subequations} and thus by the chain rule, for a general $f(g,\pi)$, if the boundary variation has been canceled by suitable boundary terms:
\be\label{equ:conf_relation} \{\pi_\phi, t_\phi f(g,\pi)\}=t_\phi\{4\pi, f(g,\pi)\}
\ee

\subsection{Canonical transformations and boundary terms}

We will here present some examples of calculating the Poisson brackets in both sides of the identity 
\be
 \{t_\phi f_1(g,\pi), t_\phi f_2(g,\pi)\}=t_\phi\{f_1(g,\pi),  f_2(g,\pi)\}
\ee
It turns out that in some cases it is easier to perform the calculation in the lhs and in others in the rhs.

\subsection{$\{t_\phi S, t_\phi S\}$ term. }
Let us start with
\be t_\phi \{ S(N_1),  S(N_2)\}= t_\phi\int d^3 x\pi^{cd}\left(N_1N_{2;cd}-N_2N_{1;cd}\right)=
\int d^3 x e^{-4\phi}\pi^{cd}\left[N_1(N_{2;cd}+4N_{1,c}\phi_{,d})-1\leftrightarrow2)\right]
\ee
where we have already set $\pi=0$ at the end result. Using integration by parts in the terms with two derivatives,  we get a derivative of $\phi $ term $e^{-4\phi}\pi^{cd}(4N_{1,c}\phi_{,d})-1\leftrightarrow2)$, which exactly cancels with the $\phi$-derivative terms above. Thus we obtain in the end 
\be\label{equ:tSS_algebra}
t_\phi \{ S(N_1),  S(N_2)\}= \int d^3 x e^{-4\phi}\pi^{cd}\left(N_1N_{2;cd}-N_2N_{1;cd}\right)+\int_{\partial\Sigma}
d^2 y e^{-4\phi}\pi^{cd}\left(N_1N_{2,c}-N_2N_{1,c}\right)
\ee
and indeed we have checked that 
\be\label{equ:tSS_bdary}
t_\phi Q(N_1,N_2):= \int_{\partial\Sigma}d^2 y r_d e^{-4\phi}\pi^{cd}(N_1N_{2,c}-N_2N_{1,c})
\ee
One can also explicitly check that indeed $\{t_\phi S, t_\phi S\}=t_\phi \{S, S\}$, although the lhs requires more algebra.

\subsection{$\{t_\phi H_ a, t_\phi H_ a\}$ term. }
For this bracket, it is easier to first notice that 
\be\label{equ:LT_H} t_\phi H_a(\xi^a)=\int d^3 x \left (\pi^{ab}\mathcal{L}_{\vec\xi}g_{ab}+4\pi\mathcal{L}_{\vec\xi}\phi\right) 
\ee
But from equation \eqref{equ:conf_relation}, for a bracket of the form $\{t_\phi H_ a, t_\phi f\}$, we can replace $4\pi\rightarrow\pi_\phi$. This simplifies the bracket and we trivially get: 
\be\int_\Sigma (\mathcal{L}_{\vec\xi_1} g_{ab}\mathcal{L}_{\vec\xi_2}\pi^{ab}+\mathcal{L}_{\vec\xi_1} \phi\mathcal{L}_{\vec\xi_2}\pi_\phi- 1\leftrightarrow 2)d^3x\ee
which gives us an extra boundary term to before: 
\be\label{equ:tHH_bdary}
t_\phi Q(\vec\xi_1,\vec\xi_2)=Q(\vec\xi_1,\vec\xi_2)+\int_{\partial\Sigma}\pi_\phi\left(\xi^a_2\mathcal{L}_{\vec\xi_1}\phi-1\leftrightarrow2\right)r_ad^2y
\ee
but upon reduction $\pi_\phi=0$, this term vanishes. Again, it is possible, and even elucidating,  to show this equality without using either the substitution  $\pi_\phi=4\pi$ or the canonical transformation identity.

\subsection{Boundary deformation algebra for Shape Dynamics}

From the constraint system $t_\phi S, t_\phi H_a, \pi_\phi-4\pi$ of the Linking theory, we will do a phase space reduction $t_{\phi_o} S=0=\pi_\phi$, and $t_\phi H_a\rightarrow H_a$. The leftover constraint $4\pi$ forms an abelian algebra with itself and thus poses no extra boundary terms.   

 To see that indeed upon reduction the diffeomorphism constraint on extended phase space goes back to the diffeomorphism constraint in original phase space, see equation \eqref{equ:LT_H}. Furthermore, from equation \eqref{equ:tHH_bdary} we know that the extra terms of the boundary deformation algebra also vanish on-shell (when $\pi=0$).  Thus we get on-shell exactly the same term as the usual \eqref{equ:HH_bdary}, and off shell we get a modification: 
\be \delta Q(\vec\xi_ 1,\vec\xi_2)=\int_{\partial\Sigma}4\pi\left(\xi^a_2\mathcal{L}_{\vec\xi_1}\phi_o-1\leftrightarrow2\right)r_ad^2y\approx 0\ee
where $\phi_o$ is the solution to the Lichnerowiz equation with a given boundary condition for $\phi$. 
 
The only potentially different leftover term that we might have gotten to the boundary deformation algebra would come from the mixed term $\{\pi, H_a\}$. A straightforward calculation yields
\be \{\pi (\rho), H_a(\xi^a)\}=\int_\Sigma d^3 x \xi^c \rho_{,c} \pi+\int _{\partial\Sigma}\xi^c\pi\rho r_c d^2 y
\ee  which also vanishes on-shell. 

So gathering these terms, we have for the total boundary deformation algebra: 
\be Q[(\vec\xi_1, \rho_1), (\vec\xi_2, \rho_ 2)]= \int_{\partial\Sigma} d^2 y \left(\mathcal{L}_{\vec\xi_1}(g_{ab})\xi^c_2 \pi^{ab}-\mathcal{L}_{\vec\xi_2}(g_{ab})\xi^c_1 \pi^{ab}+  [\vec\xi_ 1, \vec\xi_ 2]_a\pi^{ac}+\pi(\rho_1\xi_2^c-\rho_2\xi_1^c)\right)r_c
\ee





\end{appendix}

\section*{Acknowledgments} 
I would like to thank  Tim Koslowski for discussions and Steve Carlip for discussions and a careful reading of the draft. 
 HG was supported in part by the U.S.
Department of Energy under grant DE-FG02-91ER40674.

\end{document}